\documentclass[iop,numberedappendix]{emulateapj}
\usepackage{amssymb}
\usepackage{times}
\usepackage{epsfig}
\usepackage{graphics}
\usepackage{epstopdf}
\usepackage{amsmath}
\usepackage{color}
\usepackage[normalem]{ulem}

\def\rmd{{\rm d}}

\def\Te{T_{\rm e}}

\def\me{m_{\rm e}}
\def\sigmaT{\sigma_{\rm T}}
\def\tauT{\tau_{\rm T}}

\def\gff{\overline{g}_{\rm ff}}
\def\afs{\alpha_{\rm f}}
\def\tdyn{t_{\rm dyn}}

\def\rmax{r_{\rm max}}

\def\mp{m_{\rm p}}

\def\dotN{\dot{\cal N}}
\def\dotNinj{\dot{N}_{\rm inj}}

\def\rmd{{\rm d}}

\def\Te{T_{\rm e}}

\def\me{m_{\rm e}}
\def\mp{m_{\rm p}}
\def\sigmaT{\sigma_{\rm T}}
\def\tauT{\tau_{\rm T}}

\def\gff{g_{\rm ff}}
\def\afs{\alpha_{\rm f}}
\def\tdyn{t_{\rm dyn}}

\def\rmin{r_{\rm min}}
\def\rmax{r_{\rm max}}

\def\tauffep{\tau_{\rm ff,\pm}}

\def\ndense{n_{\rm dense}}
\def\kB{k_{\rm B}}
\def\lambdaC{\lambda_{\rm C}}

\def\Npl{{\cal N}_{\rm BB}}
\def\epsBB{\varepsilon_{\rm BB}}
\def\epsB{\varepsilon_B}

\def\epsinj{\varepsilon_{\rm inj}}
\def\epsrad{\varepsilon_{\rm rad}}
\def\dr{\Delta r}

\def\gdc{g_{\rm DC}}

\def\xB{x_{B}}
\def\ginj{\gamma_{\rm acc}}
\def\Gmax{\Gamma_{\rm max}}
\def\Qinj{Q_{\rm inj}}
\def\Urad{U_{\rm rad}}
\def\UB{U_{B}}

\def\jsyn{j_{\rm synch}}

\def\rmax{r_{\rm max}}
\def\Epk{E_{\rm pk}}

\def\Ph{P_{\rm h}}
\def\je{j_{\rm e}}
\def\alphae{\alpha_{\rm e}}
\def\De{D_{\rm e}}

\def\LE{L_{\rm E}}
\def\Epk{E_{\rm pk}}

\def\Lt{L_{\rm t}}
\def\rTh{r_{\star}}

\newcommand{\eqb}{\begin{equation}}
\newcommand{\eqe}{\end{equation}}

\begin{document}

\title{On thermalization in gamma-ray burst jets and the peak energies of photospheric spectra}

\shorttitle{ON THERMALIZATION IN GRB JETS AND THE PEAK ENERGIES OF PHOTOSPHERIC SPECTRA}
\shortauthors{VURM, LYUBARSKY, \& PIRAN}

\author{Indrek Vurm,\altaffilmark{1,2} Yuri Lyubarsky\altaffilmark{3} and Tsvi Piran\altaffilmark{1}}

\affil{$^1$Racah Institute of Physics, Hebrew University of Jerusalem,
Jerusalem 91904, Israel; indrek.vurm@gmail.com \\
$^2$Tartu Observatory, T\~{o}ravere 61602, Tartumaa, Estonia\\
$^3$ Physics Department, Ben-Gurion University, P.O. Box 653, Beer-Sheva 84105, Israel \\
}


\begin{abstract}
\noindent
The low energy spectral slopes of the prompt emission of most gamma-ray bursts (GRBs)
are difficult to reconcile with radiatively efficient optically thin emission models irrespective
of the radiation mechanism. An alternative is to ascribe the radiation around
the spectral peak to a thermalization process occurring well inside the Thomson photosphere.
This quasi-thermal spectrum can evolve into the observed non-thermal shape by additional energy release
at moderate to small Thomson optical depths, which can readily give rise to the hard spectral tail.
The position of the spectral peak is determined by the temperature and
Lorentz factor of the flow in the termalization zone,
where the total number of photons carried by the jet is established.
To reach thermalization, dissipation alone is not sufficient and photon generation requires
an efficient emission/absorption process in addition to scattering.
We perform a systematic study of all relevant photon production mechanisms
searching for possible conditions in which thermalization can take place.
We find that a significant fraction of the available energy should be dissipated at intermediate radii,
$\sim 10^{10}$-- a few$\times 10^{11}$ cm
and the flow there should be
relatively slow: the bulk Lorentz factor could not exceed a few tens
for all but the most luminous bursts with the highest $\Epk$-s.
The least restrictive constraint for successful thermalization, $\Gamma\lesssim 20$,
is obtained if synchrotron emission acts as the photon source. This requires, however,
a non-thermal acceleration deep below the Thomson photosphere transferring
a significant fraction of the flow energy to relativistic electrons with Lorentz factors
between 10 and 100.
Other processes require bulk flow Lorentz factors of order of a few for typical bursts.
We examine the implications of these results to different GRB photospheric emission models.

\end{abstract}


\section{Introduction}	\label{chapter:intro}

The typical broadband spectra of gamma-ray burst (GRB) prompt emission can
be broadly characterized by the following properties:
a peak at a few $100$ keV, a non-thermal power-law extending from the peak to
higher (in some cases GeV) energies, and a rising but non-thermal slope below the peak.

The enormous luminosity coupled with the non-thermal appearance of the spectrum
naturally suggests that the emission is produced by non-thermal
high-energy particles radiating in an optically thin environment.
However, optically thin emission models face serious difficulties in explaining the
{\rm low-energy turnover in} the observed spectra, regardless of
the emission process \citep{Cohen+97,Crider97,Preece98,Ghirlanda03}.
This has led to models in which energy dissipation takes place deep inside
the Thomson photosphere, where the thermalization processes are
efficient enough to produce a close to blackbody spectrum
\citep{Eichler00,MeszarosRees2000,ReesMeszaros05,Thompson07,B10}.
In such models, the radiation field, which has a quasi-thermal appearance well below the photosphere,
is distorted into a non-thermal shape by additional heating and Compton scattering
near the optically thick-thin transition.

An important question in photospheric GRB emission models concerns
the origin and amount of photons present in the flow at the stage when it approaches transparency.
In the simple fireball model these photons are created near the central engine
and advected with the jet towards the photosphere. The typical energy available per photon
at the launching site of the jet is $\sim 5$ MeV. In a pure pair-photon fireball \citep{Goodman86,Pacz86}
this would also be the average observed photon energy. On the other hand, a significant baryon load
can degrade the mean photon energy by converting most of the internal energy to bulk motion \citep{ShemiPiran90}.
However, if no more photons are created in the jet along its way, any process making the flow
radiatively efficient would bring the average photon energy back to a few MeV.
The disagreement with the observed spectral peak positions thus requires
an additional dissipation/photon production far from the center.

The frequently used assumption that any dissipation taking place well below
the Thomson photosphere leads to a thermal (Planck-like) radiation spectrum is generally invalid.
The crucial ingredient is the existence of an emission/absorption process allowing
the radiation field to attain a thermodynamic (rather than just kinetic) equilibrium with matter \citep{Beloborodov12}.
Without the creation of additional photons (as in Compton scattering)
sub-photospheric dissipation can only redistribute energy between bulk motion/Poynting flux
and the existing photons, while the total flow energy {\it per photon} remains unchanged.
Unless the radiative efficiency is low, the spectra peak at too high energies compared with observations.
The peak energy can only be decreased by introducing new photons. The lowest attainable energy
is set by the blackbody limit reached upon complete thermalization \citep{Eichler94,Eichler00,Thompson07}.

The purpose of this work is to study the efficiency of photon production and thermalization
deep inside the photosphere of GRB outflows as well as the resulting spectral peak energies,
accounting for all plausible emission/absorption processes.
For GRB jets the possible candidates are
cyclo-synchrotron emission, bremsstrahlung and double Compton scattering.
Subject to the existence of a photon source,
the relaxation of the radiation field to a Planck spectrum can be achieved in two ways:
the most straightforward option is for the jet to be optically thick to absorption throughout the spectrum.
The second and less restrictive option can be realized if the photons provided by the emission process
are redistributed into a thermal spectrum by saturated Comptonization.
In the latter case two conditions have to be satisfied:
the number of produced photons must be sufficient to fill the Planck spectrum,
and the Compton $y$-parameter has to be large for Comptonization to proceed in a saturated regime.

We examine the problem both analytically and numerically.
For the latter we use the kinetic code developed by \citet{VP09} (see also \citealt{VBP11}) that
self-consistently solves the coupled kinetic equations for electrons
and photons in relativistic flows. For the present problem the code was modified
to include induced Compton scattering, which was missing in the original version.

The paper is organized as follows: We begin by briefly outlining the spectral
hardness problem of optically thin emission models. We continue with a general
discussion of photon production and thermalization in the jet as well as their
relation to the observed spectral peak energies in the context of optically thick emission models.
This is followed by a detailed study of various processes acting as photon sources in GRB jets.
Finally, we discuss the results and their implications to GRB models.

\section{The low-energy turnover problem}

\label{section:turnover}

The prompt emission from GRBs is observed in a wide frequency band,
from hard X-rays to hard gamma-rays. The energy spectrum, $E^2dN_{\gamma}/dE$,
peaks at $\Epk\sim 100-1000$ keV. The observed spectral peak-luminosity relation is
\citep{Yonetoku04,Nava08,GhirlandaNava09,GhirlandaNava10,Gruber11}
\eqb
\Epk=300L^{1/2}_{{\rm rad},52}\,\rm keV,
\label{Amati}\eqe
where $L_{\rm rad}$ is the observed isotropic equivalent luminosity,
$\Epk$ is the spectral peak energy in the local rest frame of the burst
and the notation $A = 10^x A_x$ in cgs units has been used.
Above the peak, the hard tail with the photon
index a bit steeper than $\alpha=-2$ extends till hundreds of MeV (and even a few GeV in some cases),
which implies a non-thermal emission mechanism, e.g. synchrotron or inverse Compton.

The distribution of photon indices below the peak is centered at $\alpha\approx -1$ \citep[e.g.][]{Kaneko06,Goldstein12}
and extends to values as hard as $\alpha=1$ or even beyond \citep{Ghirlanda03}.
The hard low-energy spectra pose a great challenge to models in which the whole spectrum
is produced by relativistic electrons radiating in the optically thin regime.
The synchrotron line of death, $\alpha>-2/3$ \citep{Katz94,Cohen+97},
is violated by significant fraction of bursts \citep{Preece98}, which thus defy
explanation by optically thin synchrotron emission. Furthermore, high radiative efficiency
in the prompt phase implies that electrons emit in the fast cooling regime.
Independently of the emission mechanism, this leads to strong excess
in the low-energy part of the spectrum, in apparent contradiction with observations.
This is the low-energy turnover problem \citep[see also][]{ImamuraEpstein87}.

The problem can be understood in general terms by considering any radiation mechanism
in which the typical energy, $E$, of the emitted photon scales with the electron's Lorentz factor,
$\gamma$, as $E\propto\gamma^\rho$. An electron cooling by $\Delta\gamma$ emits
the energy $f_{\rm E} \Delta E = \me c^2\Delta\gamma$, where $f_{\rm E}$ is
the fluence per photon energy interval. The fluence due to a single electron cooling
all the way down from its initial energy is then
$f_{\rm E} = \me c^2 \rmd\gamma/\rmd E \propto E^{1/\rho - 1}$.
Since the cooling time is short, all accelerated electrons will cool.
The emissivity at a given energy $E$ is proportional to the rate at which
electrons are injected above the Lorentz factor $\gamma$ corresponding to $E$,
which equals to the flux of electrons through $\gamma$ towards lower energies.
If the injection rate is $\dotNinj\propto \gamma^{-p}$, this flux is
$\propto \gamma^{-p+1}\propto E^{(-p+1)/\rho}$ if $p>1$, and it is constant
for harder injection spectra ($p<1$). Correspondingly, the emissivity is
$j_{\rm E}\propto f_{\rm E} E^{(-p+1)/\rho}\propto E^{(2-p-\rho)/\rho}$
or $j_{\rm E}\propto E^{1/\rho - 1}$ in the cases $p>1$ and $p<1$, respectively.
Note that except for the requirement of fast cooling, the above argument is independent
of the interaction cross-section or electron cooling rate. Assuming hard electron injection ($p<1$),
the hardest spectra that can be expected from synchrotron or Compton scattering
in the Thomson regime ($\rho=2$) have a photon index $\alpha = 1/\rho - 2 = -1.5$,
whereas for Compton in the Klein-Nishina regime and bremsstrahlung ($\rho=1$) one finds $\alpha = -1$.
Therefore spectra harder than $\alpha > -1$ are very difficult to produce by
optically thin emission\footnote{Somewhat harder spectra can be produced
if {\it different} mechanisms are responsible for electron cooling and its emission
in the spectral range of interest (see e.g. \citealt{Daigne11}). A specific mechanism
that can keep accelerating the same (sub-)population of electrons can also lead to harder spectra:
in this case the limit $\alpha=-2/3$ could in principle be achieved for synchrotron emission.
For inverse Compton the corresponding limit is $\alpha=0$ (see e.g. \citealt{SP04} for a particular model)}.

\section{The photon production problem}

\label{section:phprod}

The low-energy turnover problem is avoided if the low-energy spectrum is shaped in
the optically thick regions of the flow. Such photospheric emission models usually invoke
some form of dissipation close to the Thomson photosphere to account for
the non-thermal appearance of the overall spectrum
\citep[e.g.][]{ReesMeszaros05,Peer2006,Giannios06,GianniosSpruit07,Giannios2008,B10,VBP11}.
However, as shown below, the total {\it number} of photons carried by the jet is established
at much smaller radii. Along with the luminosity this number determines the position of the spectral peak,
compatibility with the observed $\Epk$ values thus requires sufficiently efficient photon production in the flow.

For a given luminosity, radius and Lorentz factor the lowest $\Epk$ value is attained
if the spectrum is a blackbody. Consider the case where a fraction $\epsBB$ of
the total available energy $L$ is dissipated and processed into a Planck spectrum
at radius $r$, which we call the thermalization radius.
The isotropic equivalent radiation luminosity at that location is given by
\eqb
\epsBB L=4\pi cr^2\Gamma^2\frac 43aT^4,
\label{bb}
\eqe
where $\Gamma$ and $T$ are are the jet Lorentz factor and (comoving) temperature at $r$, respectively.

Between the thermalization radius and radius where the radiation decouples from
the flow (i.e. the Thomson photosphere), the radiation luminosity can be altered
by adiabatic cooling as well as further dissipation. Let's define $\epsrad$ as
the final radiative efficiency, i.e. the fraction of the total energy $L$ released to
the observer ($L_{rad} \equiv \epsrad L$). Without further dissipation beyond
the thermalization zone, we have $\epsrad\le \epsBB$, where the equality holds
if the flow remains radiation-dominated until the Thomson photosphere.
The inequality holds in a coasting flow, in which case the average photon energy
is degraded by adiabatic cooling. With further dissipation one may also have $\epsrad> \epsBB$.

If no more photons are produced beyond the thermalization zone,
the observed spectral peak is approximately given by
\eqb
\Epk=6\kB T \, \Gamma \, \frac{\epsrad}{\epsBB},
\label{thermal_peak}\eqe
where $6\kB T \, \Gamma$ is the approximate $\nu F_{\nu}$ peak energy of
a Lorentz boosted blackbody spectrum\footnote{The $\nu F_{\nu}$ peak of
a blackbody spectrum in the comoving frame is at $3.92\kB T$,
which is boosted on average by $1.5\Gamma$ in a relativistic quasi-spherical flow,
yielding a peak at $5.83\kB T\Gamma$ in the external frame.} (see e.g. \citealt{Grimsrud98,LiSari08})
and the factor $\epsrad/\epsBB$ accounts for
adiabatic cooling and/or additional dissipation between the thermalization zone and
the Thomson photosphere. From equations (\ref{bb}) and (\ref{thermal_peak}) we obtain
\eqb
\Epk = 200\sqrt{\frac{\Gamma_2}{r_{12}}} \, \frac{\epsrad}{\epsBB} \left(\epsBB L_{52}\right)^{1/4}\,\rm keV.
\label{thermal_peak2}
\eqe
Comparing with the observed position of the spectral peak (\ref{Amati})
and using $L_{\rm rad}\equiv\epsrad L$, one finds a relation between the thermalization radius and
the flow Lorentz factor in the thermalization zone \citep{Eichler00,Thompson07}:
\eqb
\frac{r}{\Gamma}=5\cdot 10^9 \, \frac{\epsrad}{\epsBB^{3/2}}  \, L_{52}^{-1/2}\,\rm cm.
\label{r-gamma}
\eqe
An immediate consequence of Equation (\ref{r-gamma}) is that
the observed positions of the spectral peaks cannot be explained by relic photons
produced near the central engine, unless the radiative efficiency is very low.
Instead, most of the observed photons have to be produced far from the center.

The derivation of Equation (\ref{r-gamma}) assumes that the flow maintains
a constant opening angle between the thermalization location and the Thomson photosphere.
If the flow undergoes substantial collimation between these regions, the constraint
on $r/\Gamma$ is somewhat relaxed (Equation (\ref{r-gamma:coll}) in
Appendix \ref{section:app:collim}; see also \citealt{Beloborodov12}). However, for realistic jet opening angles
the constraint on $r$ is still incompatible with the size of the central engine,
thus our conclusion that bulk of the observed photons have to be produced
further in the jet still holds.

Collimation that operates after the thermalization zone would somewhat
relax condition (\ref{r-gamma}). However, most of the jet collimation is likely
to take place below the thermalization zone. We will therefore use
Equation (\ref{r-gamma}) as it stands and ignore the effects of possible
further collimation at larger radii, which could influence the constraint
on $r/\Gamma$ at most by a factor of a few.

There are several aspects to be considered regarding the way thermalization
can be achieved in GRB jets. First, one has to stress that thermalization requires that
blackbody radiation is formed. For that the thermalization zone should be optically thick to
absorption. Dissipation alone below the Thomson photosphere is {\it insufficient}.
The photon production problem arises because the rates of emission/absorption processes
rapidly decrease with the distance and with the Lorentz factor of the flow
(because all relevant parameters like the proper density and the magnetic field decrease).
Therefore, it is difficult to find a powerful enough photon source and satisfy
Equation (\ref{r-gamma}) simultaneously. This combination requires rather specific
conditions within in the jet, as we show below.

Secondly, thermal-like spectra can be formed
via Comptonization of soft photons on thermal electrons
\citep[e.g.][]{Liang97,GhiselliniCelotti99,Thompson07,Giannios12,Beloborodov12}.
In this case, the spectrum also peaks at the energy (\ref{thermal_peak}),
however the energy density may be less than $aT^4$,
therefore Equation (\ref{r-gamma}) becomes an inequality so that
the production of observed photons should take place
at an even larger distance.
On the other hand, if the emission/absorption process is able to provide enough photons,
complete thermalization may be achieved and Equation (\ref{r-gamma}) remains as it stands.
In general, the conditions are less restrictive for thermalization by Comptonizing soft photons
than by the absorption process alone.

The Comptonization efficiency also places limits on the parameters in the thermalization zone.
The condition that photons are efficiently redistributed towards the thermal peak is
\eqb
y = 4 \, \frac{\kB T}{\me c^2} \, \sigmaT N c\tdyn \gtrsim 10,
 \label{Compton}\eqe
where $y$ is the Compton parameter,
$N$ is the proper electron number density and $\tdyn=r/c\Gamma$ the proper propagation/dynamical time.
One can conveniently normalize the proper plasma density by the Lorentz factor corresponding to
the total energy conversion into the kinetic energy, $\Gmax$:
 \eqb
 4\pi r^2c N\Gamma=L/(\mp c^2\Gmax).
\label{eq:prob:N}
 \eqe
Then the Thomson optical depth of the flow is \citep[e.g][]{Abramowicz91}
 \eqb
\tauT=\sigmaT Nc \tdyn= \frac{\sigmaT Nr}{\Gamma}
=1.2\cdot 10^4\frac{L_{52}}{r_{10}\Gamma_1^2\Gamma_{\rm max,3}}.
\label{eq:prob:tau}
\eqe
 The plasma temperature can be expressed via the
peak energy ($\ref{thermal_peak}$).
Using the observed $\Epk$ -- $L_{\rm rad}$ relation (\ref{Amati}),
the condition (\ref{Compton}) can now be written as
\eqb
100\frac{\epsBB L_{52}^{3/2}}{\epsrad^{1/2} \, r_{10}\Gamma_1^3\Gamma_{\rm max,3}}\gtrsim 1.
\label{y>1}\eqe
Taking into account condition (\ref{r-gamma}),
this constrains the Lorentz factor in the thermalization zone to a few tens:
\begin{align}
\Gamma \lesssim 21 \frac{\epsBB^{5/8} \, L_{52}^{1/2}}{\epsrad^{3/8} \, \Gamma_{\rm max, 3}^{1/4}}.
\label{saturated}\end{align}
Thus, the kinetic energy should be small in that location.
We note in passing that this could suggest that the energy is transferred
from the central engine to the thermalization zone in the form of the Poynting flux
which accelerates much slower than a thermal fireball. Alternatively, one has to find
a highly efficient mechanism to convert almost all of the available kinetic energy
to internal form at $r\sim 10^{10}$ -- $10^{11}$ cm.

Even stronger restrictions are obtained from the consideration of
the soft photon sources necessary for the Comptonization. The condition (\ref{y>1})
only ensures that the photons are redistributed towards the observed peak energy.
However, one has to find a photon source capable of producing
the observed amount of photons. The condition for the complete thermalization is written as
\begin{align}
 \tdyn \,\dotN \geq \Npl,
\label{thermalization}\end{align}
where $\dotN$ is the photon injection rate per unit volume, and
\begin{align}
\Npl \approx  2.4 \, \frac{8\pi}{\lambdaC^3} \, \theta^3
\label{Plank}\end{align}
is the total number density of photons in the Planck spectrum.
Here $\theta=\kB T/\me c^2$ is the dimensionless temperature and
$\lambdaC=h/\me c$ is the Compton wavelength. We show below that
the requirement that this condition is satisfied simultaneously with
Equation (\ref{r-gamma}) places severe limits on the parameters of the thermalization zone.

In unmagnetized flows the only plausible emission/absorption mechanisms
are bremsstrahlung and double-Compton scattering. Their corresponding
absorption photospheres are deep below the Thomson photosphere, at $\tauT\gtrsim 10^4$.
Due to Comptonization, thermalization can occur above the absorption photosphere but not far from it.
In magnetized flows, one has also to take into account cyclotron and synchrotron emission,
which are copious sources of low energy photons. In this case saturated Comptonization
of soft photons is crucial for thermalization.

\section{The physical model}

To study the thermalization and photon production capabilities of the jet by various processes,
we consider an energy dissipation episode taking place between $\rmin$ and $\rmax=2\rmin$
in magnetized or unmagnetized outflows carrying baryons, electrons and radiation
(and at some conditions, electron-positron pairs). The condition (\ref{r-gamma}) requires
that a sizable fraction of the total available energy -- kinetic or magnetic -- be converted
to internal (thermal) energy at radii $r\gtrsim 10^{10}$ cm. Without specifying
the conversion mechanism, we assume that this energy is supplied to the electrons
either in the form of heating the bulk of the lepton population, or by accelerating a fraction of them
to relativistic energies.
The electrons transfer their energy to radiation via emission of photons by various processes
and their Comptonization.
The main questions we ask for each emission process are whether
there is time to emit and reprocess enough photons to fill
the Planck spectrum and whether the required conditions for this are compatible with
the general constraint (\ref{r-gamma}).
We also check the possibility of partial thermalization when a Wien rather than
the Planck spectrum is formed with the peak energy compatible with the observed relation (\ref{Amati}).

We study the problem analytically as well as numerically in more complicated
cases involving magnetized jets. The analytic study is based on analysis of
the Kompaneets equation, which allows us to estimate the amount of photons
available for thermalization without solving the full equation (see Appendix \ref{section:app:Compt}).
For numerical simulations we use the kinetic radiative transfer code developed
by \citet{VP09} and \citet{VBP11} (Appendix \ref{section:app:method}).

\section{Thermalization by bremsstrahlung}

\label{section:brems}

Bremsstrahlung can serve as a photon source
if the density is sufficiently large.
This can happen only if energy is dissipated deep in the flow
where the temperature is high enough to sustain a significant population of electron-positron pairs.
If the bremsstrahlung optical depth is above unity all the way to the thermal peak energy $E\sim 3\kB T$,
complete thermalization is obviously achieved.
However, a somewhat weaker condition for thermalization can be obtained by
taking into account Comptonization of bremsstrahlung photons.

Making use of Equation (\ref{Ndot_ff}) in Appendix \ref{section:app:Brems} for
the bremsstrahlung photon injection rate, one can write
the thermalization condition ({\ref{thermalization}}) as
\begin{align}
\frac{\dotN_{\rm ff} \, \tdyn}{\Npl}=
\frac{\tau_{\rm ff}^{\pm}}{4.8} \,\, \ln^2\frac{2.35\kB T}{E_0} \gtrsim 1,
\label{eq:thermcond}
\end{align}
where
\eqb
\tau_{\rm ff}^{\pm}=
\frac{\afs}{3^{1/2} \,2\pi^{3/2}}  \, \left(\frac{\kB T}{\me c^2}\right)^{-7/2} \,  \lambdaC^3 \,N_{\pm}  \tauT
\label{eq:tauff}
\eqe
is the free-free optical depth at $E=\kB T$.
The transition energy $E_0$ above which bremsstrahlung photons are
Comptonized rather than reabsorbed is given by Equation (\ref{eq:nu0}).

Condition (\ref{eq:thermcond}) can be satisfied only at the stage where
the flow is still heavily loaded with pairs, which are in (quasi-)thermodynamic
equilibrium with the radiation. If the radiation field is a blackbody,
the electron/positron chemical potential is $\mu_{\pm}=-\me c^2$
and the density of electrons/positrons is
\begin{align}
N_\pm = \frac{2}{\lambdaC^3} (2\pi\theta)^{3/2} e^{-1/\theta}.
\label{eq:nlep}
\end{align}
Substituting this together with Equation (\ref{eq:tauff}) into Equation (\ref{eq:thermcond})
yields a constraint on temperature where thermalization is still effective,
in terms of the flow's Lorentz factor and radius. For Planckian radiation we can use
the relation (\ref{bb}) to eliminate either $T$ or $r$ and obtain
a constraint for the other variable in terms of $\Gamma$, $L$ and $\epsBB$.

We find that Comptonization can extend the thermalization region
up to $\tauffep(E \! = \! 3\kB T)\sim 0.005$. However the thermalization radius
is only about $50$\% larger than the radius of the bremsstrahlung photosphere,
owing to the strong dependence of the opacity on the temperature and thus also
on $r$ via Equation (\ref{bb}). Therefore Comptonization is relevant only in
a narrow range of radii and thus has a limited role in forming the Planckian spectrum.

\begin{figure}[h]
\plotone{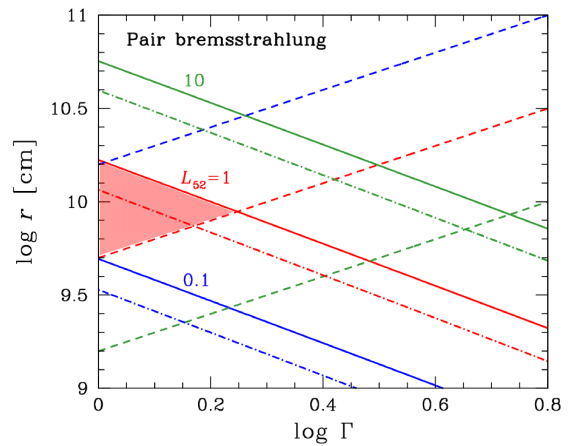}
\caption{Constraints on $r$ and $\Gamma$ from the requirement
of efficient bremsstrahlung thermalization (solid lines), and from
the observed peak energies (dashed lines), for flow luminosities
$L=10^{51}$ (blue), $10^{52}$ (red), $10^{53}$~erg~s$^{-1}$ (green) and $\epsBB=1$.
The allowed region is below the solid and above the dashed lines.
The bremsstrahlung photosphere is shown by dot-dashed lines.
}
\label{fig:r_brems}
\end{figure}

Figure \ref{fig:r_brems} depicts
the allowed parameter region for $r$ and $\Gamma$, for fixed $L$ and $\epsBB$.
The solid lines correspond to the thermalization condition (\ref{eq:thermcond})
and limit the dissipation radius and Lorentz factor from above.
The constraints from the observed peak energy from Equation (\ref{r-gamma})
are shown by dashed lines, limiting $r$ from below. The allowed parameter space
forms a wedge-shaped region to the left of the crossing point of the two constraints.
This point sets the maximal Lorentz factor in the thermalization zone for a given luminosity.
One can see that a solution exists (i.e. the maximal $\Gamma>1$) only for
luminosities above a few$\times 10^{51}$~erg~s$^{-1}$.

\begin{figure}[h]
\plotone{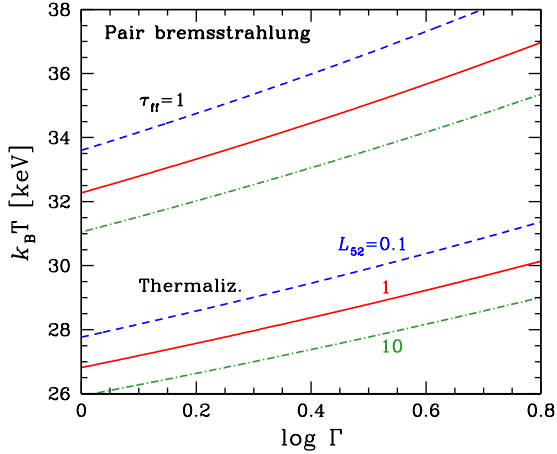}
\caption{Bremsstrahlung thermalization constraint on the comoving temperature (lower lines),
for flow luminosities $L=10^{51}$ (dashed), $10^{52}$ (solid), $10^{53}$~erg~s$^{-1}$
(dot-dashed) and $\epsBB=1$. The allowed regions for different luminosities are
above the corresponding lines. Also shown is the temperature at the bremstrahlung
photosphere (upper lines) for the same luminosities.
}
\label{fig:T_brems}
\end{figure}

The lower limits on the temperature in the thermalization zone are shown
in Figure \ref{fig:T_brems}. The temperature at which thermalization
fails depends weakly on other parameters and is around $28$ keV.
This is a consequence of the exponential dependence of the number
of pairs on $\theta$ (Equation (\ref{eq:nlep})). An approximate analytic
thermalization constraint on $r$ and $\Gamma$ can be read from Equation (\ref{bb}),
keeping $T$ constant at 28 keV:
\begin{align}
\Gamma \, r_9 \lesssim 15 \, (\epsBB L_{52})^{1/2}.
\end{align}
This constraint is compatible with the condition (\ref{r-gamma}) only if
\begin{align}
\Gamma \lesssim 2 \, \frac{\epsBB}{\epsrad^{1/2}} \, L_{52}^{1/2},
\label{eq:brems:Gamma}
\end{align}
i.e. the flow can only be mildly relativistic at the thermalization zone\footnote{For modification of this constraint for bursts deviating from the Yonetoku relation, as well as the modifications of analogous constraints for other processes discussed below, see Appendix \ref{section:app:yonet}.}.

\subsection{The effect of clumping}

In a smooth flow bremsstrahlung rapidly becomes insignificant as a source
of photons once the pairs drop out. However, by virtue of the quadratic dependence
on density the average emission/absorption rates can be increased if regions
of the flow are locally compressed to higher densities. Such situation can arise,
for example, in a Poynting dominated flow carrying an alternating magnetic field (striped wind).
This implies the existence of current sheets where matter density is substantially
enhanced compared to the average in the flow.

Let's assume that a fraction $\lambda$ of the plasma is compressed in
the radial direction by a factor $\xi=l/\Delta$ so that $\Delta$ is the average size
of the compressed region and $l$ is the average distance between such regions.
The density in the compressed regions is $N_{\rm dense}=\lambda \xi N$,
where $N$ is the average density given by Equation (\ref{eq:prob:N}).
Bremsstrahlung opacity is proportional to the emission measure
\begin{align}
<n^2> r/\Gamma \approx \ndense^2 \frac{\Delta}{l} \frac{r}{\Gamma}
= \lambda^2 \xi N^2 \frac{r}{\Gamma},
\end{align}
and is given by Equations (\ref{eq:absff1}) and (\ref{ep_ff}).
Radiation is thermal as long as
$\alpha_{\rm ff}t_{\rm dyn}\ge 1$ at $E =3\kB T$, yielding
\begin{align}
\Gamma r_{10} \le 1.7\times 10^{-3} \, (\lambda^2\xi)^{4/5}
\, \frac{L_{52}^{9/10}}{\epsBB^{7/10} \, \Gamma_{{\rm max},3}^{8/5}},
\label{eq:brems:therm2}
\end{align}
where we have used Equations (\ref{bb}), (\ref{eq:prob:N}) and (\ref{eq:prob:tau})
to express $\kB T$, $N$ and $\tauT$, respectively. We see that high
compression factors are required for compatibility with condition (\ref{r-gamma}).

In the striped wind, the maximal expected compression can be estimated by balancing the
magnetic pressure outside the current sheets with the thermal pressure inside,
$B^2/8\pi = 2N_{\rm dense} \kB T$. Defining $\epsB$ as the fraction of total energy
carried by the Poynting flux by
\begin{align}
\epsB \, L = \frac{c r^2 \Gamma^2 B^2}{2},
\label{epsB}\end{align}
we obtain
\begin{align}
\frac{N_{\rm dense}}{N} = \lambda\xi = 4.3\times 10^6
\, \frac{\epsB \, r_{10}^{1/2} \, \Gamma_{{\rm max},3}}{\Gamma_1^{1/2} (\epsBB L_{52})^{1/4}}.
\end{align}
Using Equation (\ref{eq:brems:therm2}), this gives
\begin{align}
\Gamma r_{10}^{3/7} \le 120 \, \frac{\lambda^{4/7} \, \epsB^{4/7} \, L_{52}^{1/2}}{\epsBB^{9/14} \, \Gamma_{{\rm max},3}^{4/7}}.
\end{align}
Compatibility with condition (\ref{r-gamma}) limits the Lorentz factor to
\begin{align}
\Gamma \le 35 \,  \frac{\lambda^{2/5} \, \epsB^{2/5} \, L_{52}^{1/2}}{\epsrad^{3/10} \, \Gamma_{{\rm max},3}^{2/5}}.
\end{align}
Note that this condition was obtained neglecting Comptonization because
the condition for the saturated Comptonization (\ref{saturated}) is stronger.

\section{Thermalization by double Compton scattering}

Double Compton scattering, where an additional photon is produced upon
the scattering of an electron and photon, can supply enough photons for
thermalization if the dissipation operates deep enough below the Thomson photosphere.
Making use of the corresponding photon injection rate (\ref{eq:DCinj})
(Appendix \ref{section:app:DC}), one writes the thermalization condition (\ref{thermalization}) as
\begin{align}
\frac{\dotN_{\rm DC} \, \tdyn}{\Npl}=
\tauT \, \frac{16\afs}{\pi} \, \theta^2 \gdc(\theta) \, \ln\frac{\kB T}{E_0} \ge 1,
\label{eq:DC:therm}
\end{align}
where the transition energy $E_0$ is given by Equation (\ref{eq:DC:nu0}).
The blackbody relation (\ref{bb})
can be used together with Equation (\ref{eq:prob:tau}) for $\tauT$
to eliminate one of the variables $\theta$, $r$ or $\Gamma$
from the thermalization condition (\ref{eq:DC:therm}).
Eliminating either $r$ or $\Gamma$, we find that temperature at the thermalization
limit is a weak function of the other variables, scaling approximately as
\begin{align}
\theta_{\star} \propto L^{-1/8} \Gamma^{1/4} \Gmax^{1/4} \quad \mbox{or} \quad \theta_{\star} \propto r^{-1/6} \Gmax^{1/6}
\end{align}
and has value around $5$ -- $10$ keV for typical parameters \citep[see also][]{Beloborodov12}.
Consequently, Equation (\ref{eq:DC:nu0}) implies that the logarithmic factor
in Equation (\ref{eq:DC:therm}) is approximately constant, $\ln(\kB T/E_0)\approx 4$.
The constraint itself becomes
\begin{align}
\Gamma \, r_{10}^{2/3} \lesssim 8.5 \, \frac{\epsBB^{1/6} L_{52}^{1/2}}{\Gamma_{{\rm max},3}^{1/3}}.
\label{eq:DC:thermcond}
\end{align}
Together with condition (\ref{r-gamma}) this requires
\begin{align}
\Gamma \lesssim 4.8 \, \frac{\epsBB^{7/10} L_{52}^{1/2}}{\epsrad^{2/5} \, \Gamma_{{\rm max},3}^{1/5}}.
\label{eq:DC:Gamma}
\end{align}

\begin{figure}[h]
\plotone{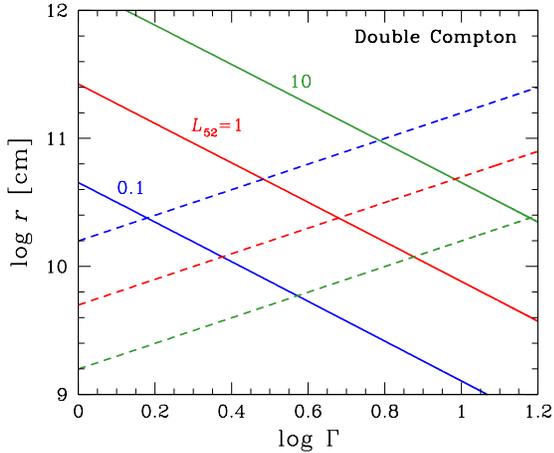}
\caption{Constraints on $r$ and $\Gamma$ from the requirement
of efficient thermalization by double Compton scattering (solid lines),
and from observed peak energies (dashed lines). The blue, red and green
lines correspond to luminosities $L=10^{51}$, $L=10^{52}$ and $L=10^{53}$~erg~s$^{-1}$, respectively.
In all cases $\epsBB=1$. For each luminosity the allowed region is below
the corresponding solid and above the corresponding dashed line.
}
\label{fig:r_DC}
\end{figure}

Figure \ref{fig:r_DC} depicts the allowed parameter space $r$ and $\Gamma$
for double Compton. The thermalization constraint (\ref{eq:DC:thermcond})
is shown by solid lines and the dashed lines correspond to the constraint from
the observed peak energies, Equation (\ref{r-gamma}). Comparison with
Figure \ref{fig:r_brems} reveals that thermalization is achieved via
the double Compton scattering a bit easier than via bremsstrahlung.
Still the conditions are rather stringent; the process works efficiently
only if the energy is released in a mildly relativistic flow.

\section{Thermalization by Comptonizing cyclotron radiation}

In highly magnetized flows, one has to take into account cyclotron radiation
of thermal electrons. If the fraction $\epsB$ of the total energy is transferred
by the Poynting flux according to Equation (\ref{epsB}), the magnetic field in
the comoving frame is
 \eqb
B=\frac 1{\Gamma}\sqrt{\frac {2\epsB L}{r^2c}}=8.2\cdot 10^7\frac{L^{1/2}_{52}}{r_{11}\Gamma_2}\, \rm G.
 \eqe
Because of strong self-absorption, most of photons are emitted at
high cyclotron harmonics, i.e. at $E\gg E_B$ where
\begin{align}
 E_B= \frac{heB}{2\pi\me c}
 = 0.95 \, \frac{\sqrt{\epsB L_{52}}}{r_{10} \Gamma} \, \mbox{keV}
\label{eq:cycl:xB}
\end{align}
is the cyclotron energy in the comoving frame.

\begin{figure}[h]
\plotone{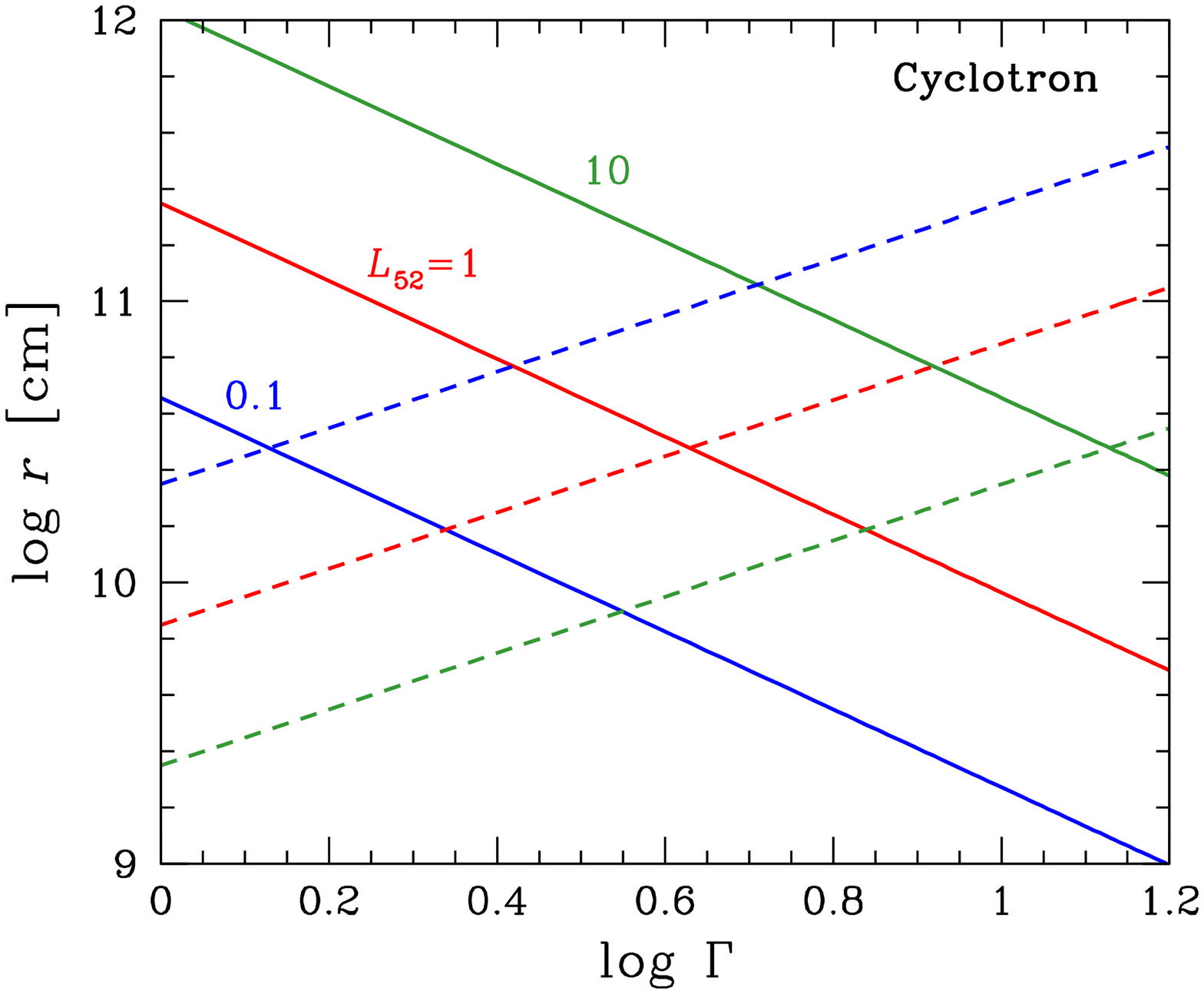}
\caption{Constraints on $r$ and $\Gamma$ from the requirement of
efficient thermalization by cyclotron emission (solid lines), and from
observed peak energy (Equation (\ref{r-gamma}), dashed lines).
The blue, red and green lines correspond to luminosities
$L=10^{51}$, $L=10^{52}$ and $L=10^{53}$~erg~s$^{-1}$, respectively.
Other parameters: $\epsrad=\epsB=\epsBB=0.5$, $\Gmax = 10^3$.
For each luminosity the efficient thermalization region is below the corresponding solid lines.
The crossing point of the solid and dashed lines defines the maximal Lorentz factor at which
the thermalization condition is compatible with condition (\ref{r-gamma}).
}
\label{fig:r_cycl}
\end{figure}

Making use of equations (\ref{Plank}) and (\ref{eq:cycl:dotN}) in Appendix \ref{section:app:cycl},
one can compare the amount of the emitted photons with that in the Planck spectrum,
\begin{align}
\frac{\dotN_{\rm cycl} \, \tdyn}{\Npl}
&= \frac{1}{1.6} \, \tauT\frac{E_0^2}{\me c^2\kB T}
= 130 \tauT \, \theta^{-2/5} \, \left(\frac{E_B}{\me c^2} \right)^{9/5}							 \nonumber \\
&=1.3 \, \frac{\epsB^{9/10} \, L_{52}^{9/5}}{\epsBB^{1/10} \, r_{10}^{13/5}\Gamma_1^{18/5}\Gamma_{\rm max,3}},
\end{align}
where in the last step we have used Equations (\ref{bb}), (\ref{eq:prob:tau})
and (\ref{eq:cycl:xB}) to express $\theta$, $\tauT$ and $\xB$ in terms of outflow parameters.
One sees that Comptonization of the thermal cyclotron radiation could provide
complete thermalization if the following condition is satisfied
\eqb
\Gamma \, r_{10}^{13/18}\lesssim 11 \, \frac{\epsB^{1/4}L_{52}^{1/2}}{\epsBB^{1/36}\, \Gamma_{\rm max,3}^{5/18}}.
\label{eq:cycl:thermcond}
\eqe
Compatibility with condition (\ref{r-gamma}) again restricts the flow Lorentz factor
to mildly relativistic values in the thermalization region,
\eqb
\Gamma \lesssim 5.4 \, \frac{\epsBB^{19/31}\epsB^{9/62}L_{52}^{1/2}}{\epsrad^{13/31}\, \Gamma_{\rm max,3}^{5/31}}.
\label{eq:cycl:Gamma}
\eqe
The constraint (\ref{eq:cycl:thermcond}) on radii and Lorentz factors of
the thermalization zone is shown in Figure \ref{fig:r_cycl} (solid lines),
along with the constraint (\ref{r-gamma}) from observed peak energies (dashed lines).

If condition (\ref{eq:cycl:thermcond}) is not satisfied and complete
thermalization is not achieved, the position of the spectral peak can be estimated
by equating the total dissipated luminosity to injection rate of cyclotron photons times
the energy they attain by saturated Comptonization\footnote{Henceforth we use $\epsBB$
to denote the fraction of total energy dissipated in the photon-production zone
regardless of whether a blackbody spectrum is attained or not,
while $\epsrad$ still refers to the radiative efficiency as seen by the observer.}:
\begin{align}
\epsBB L = \frac{16\pi}{3} r^2 \dr \,  3 \kB T\, \Gamma \, \dotN_{\rm cycl},
\end{align}
where $\dr$ is range over which the dissipation operates and the
Lorentz invariant $\dotN$ is given by equation (\ref{eq:cycl:dotN}).
We find
\begin{align}
\Epk \approx 6\Gamma \, \kB T \, \frac{\epsrad}{\epsBB}
= 590 \, \frac{\epsrad \, r_{10}^{1/2} \, \Gamma_1^{3/2} \,
\Gamma_{{\rm max},3}^{5/18}}{\epsBB^{13/18} \,  \Delta r_{10}^{5/18} (\epsB L_{52})^{1/4}} \, \mathrm{keV}.
\label{eq:cycl:Epk}
\end{align}
If we take $\dr\propto r$, we obtain $\Epk\propto r^{2/9}$, i.e. the peak energy of
the spectrum increases if the dissipation zone is moved beyond the thermalization radius.

\begin{figure}[h]
\plotone{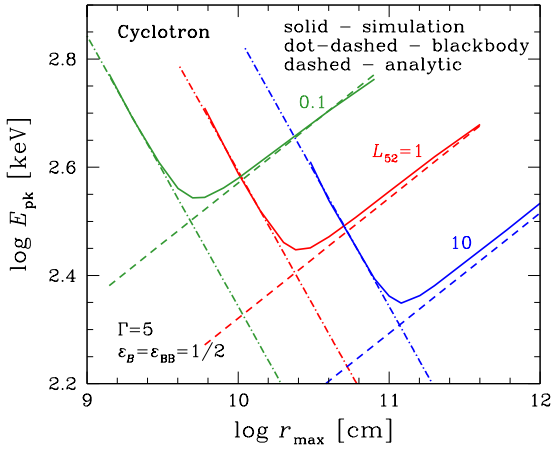}
\caption{The spectral peak energy as a function of the dissipation radius for
complete and incomplete thermalization of cyclotron radiation.
The dissipation operates in the range $[\rmax/2,\rmax]$. The solid lines are
the numerical results, the dot-dashed lines correspond to the blackbody result (\ref{thermal_peak2}),
with $\epsrad=\epsBB=1/2$. The dashed lines correspond to the analytic result
for incomplete thermalization, Equation (\ref{eq:cycl:Epk}).
}
\label{fig:Epk_cycl}
\end{figure}

The behaviour of the spectral peak as a function of the dissipation radius
is shown in Figure \ref{fig:Epk_cycl} (solid lines). At sufficiently small radii,
thermalization is complete and the peak energy follows the blackbody law (dot-dashed lines).
It has a minimum at the radius where thermalization just begins to fail,
after which the spectral peak starts increasing with radius, in agreement
with the analytical result (\ref{eq:cycl:Epk}) (dashed lines).

\begin{figure}[h]
\plotone{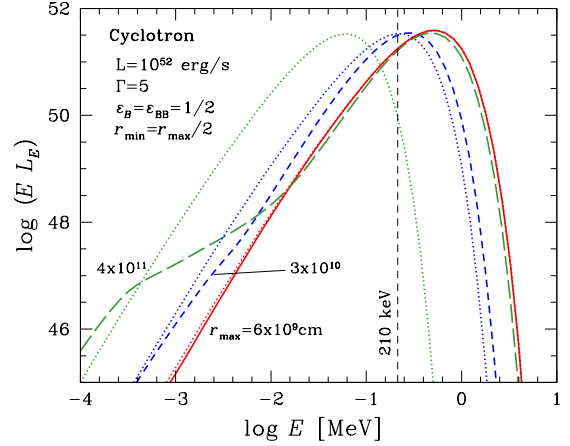}
\caption{Spectra at the end of the dissipation episode, taking place in
the range $[r_{\rm max}/2,r_{\rm max}]$. The solid (red), short-dashed (blue)
and long-dashed (green) lines correspond to different dissipation radii.
The dotted lines are the corresponding Planck spectra for the same
luminosities and radii. The vertical dashed line corresponds to the average
observed peak energy $210$ keV for $L_{{\rm rad},52}=\epsrad L_{52}=1/2$.
}
\label{fig:Spec_cycl}
\end{figure}

Figure \ref{fig:Spec_cycl} depicts the spectra at the end of the dissipation episode,
for different dissipation radii. For comparison, the corresponding thermal spectra
of the same luminosities are also shown. As expected, at small radii
the numerical result is practically indistinguishable from a Planck spectrum.
At a larger radius where thermalization fails, two distinct components appear
in the spectrum (green dashed line): the self-absorbed cyclotron emission
(below $\sim 1$ keV), i.e. Rayleigh-Jeans with brightness temperature equal to
the electron temperature, and a Comptonized spectrum at higher energies,
with $\LE\propto E^0$ at $E\ll\kB T$ and a Wien bump near $\kB T$ as expected
in a saturated regime. Note that as these spectra are produced deep in
the flow where $\tauT\gg 1$, their shape has no direct bearing on the observed
emission released at the Thomson photosphere. However, if the flow is
radiatively efficient at $\tauT\approx 1$, their peak energy is reasonably close to the final value.

\section{Thermalization by Comptonizing synchrotron radiation}

\label{section:synch}

Synchrotron emission is a copious source of photons. \citet{Thompson07}
suggested that thermalization occurs via Comptonization of synchrotron photons
when the jet breaks out of the Wolf-Rayet core at the distance of about
$10^{10}$ cm from the source. With this and some other assumptions,
they have managed to reproduce the observed relation between the peak energy
and isotropic luminosity (Equation (\ref{Amati})). When analyzing
the thermalization process, they assumed that all synchrotron photons emitted above
the synchrotron self-absorption frequency are redistributed to $E\sim 3\kB T$; then
even a small population of relativistic electrons in the presence of
a moderate magnetic field suffices for thermalization. However, most of
synchrotron photons are emitted near the self-absorption frequency
where the brightness temperature is relativistic (it is equal to the energy of emitting electrons)
whereas thermal electrons have a non-relativistic temperature. Hence the energy
could not be transferred from the "cold" electrons to the hot radiation.
On the contrary, the high brightness temperature photons lose energy
(via induced scattering, see Appendix \ref{section:app:Compt}). Only photons emitted well above
the self-absorption frequency, where the brightness temperature is
less than the temperature of thermal electrons, could be redistributed
by the Comptonization to the thermal peak. Taking account of this,
let us address the thermalization of synchrotron radiation at GRB conditions.

The synchrotron emissivity of a relativistic power-law distribution
$N(\gamma)=N_0 \gamma^{-p}$ of electrons is approximately given by
\begin{align}
&\jsyn(E) = j_{\star} \left( \frac{E}{E_B} \right)^{-(p+5)/2} \nonumber \\
&=\frac{\pi}{4} \afs^{-1} c\sigmaT N_0
\, \left(\frac{\me c^2}{E_B}\right)^2 \left( \frac{E}{E_B} \right)^{-(p+5)/2},
\label{eq:Synch:PL}
\end{align}
which gives the variation rate of the photon occupation number per unit time.
Assuming that electrons are injected with typical Lorentz factor $\ginj$
with power $\Qinj$ (per unit volume, in $\me c^2$ units), we can determine
the electron distribution function from the equation for electron cooling
\begin{align}
\frac{\partial}{\partial\gamma}\left[ \dot{\gamma} N(\gamma) \right]
+ \frac{\Qinj}{\ginj} \, \delta(\gamma-\ginj) = 0,
\label{eq:Synch:elcool}
\end{align}
where
\begin{align}
\dot{\gamma} = \frac{4}{3} \frac{\sigmaT\gamma^2 (\UB + \Urad)}{\me c},
\end{align}
where $\UB$ and $\Urad$ are the magnetic and radiation
energy densities, respectively. The solution is
\begin{align}
N(\gamma) = N_0 \, \gamma^{-2} = \frac{3\me c \,\Qinj}{4\sigmaT \, \ginj \gamma^2 \, (\UB+\Urad)}.
\label{eq:synch:Ngamma}
\end{align}
Introducing $\epsinj$ as the fraction of total energy provided to
accelerated particles over a dynamical time, we can write
\begin{align}
\Qinj = \frac{\epsinj L}{4\pi c r^2\Gamma^2}.
\end{align}
Using this in Equation (\ref{eq:synch:Ngamma}) and substituting $N_0$
back to the emissivity expression (\ref{eq:Synch:PL}), we obtain
\begin{align}
j_{\star} = \frac{3\pi}{16} \afs^{-1} \, \frac{c\Gamma}{r} \,
\frac{\epsinj}{\ginj (\epsB+\epsBB)} \left(\frac{\me c^2}{E_B}\right)^2.
\label{eq:synch:j0}
\end{align}
The distribution (\ref{eq:synch:Ngamma}) also yields $p=2$ in Equation (\ref{eq:Synch:PL}).

The transition energy $E_0$ above which photons are upscattered is found from
condition (\ref{eq:synch:upsc}) in Appendix \ref{section:app:synch}, substituting $\jsyn$ from
Equations (\ref{eq:Synch:PL}) and (\ref{eq:synch:j0}),
\begin{align}
\left(\frac{E_0}{E_B}\right)^{5/2} =\frac{32}{81} \frac{\me c}{\sigmaT N} \frac{E_B}{(\kB T)^2}  \, j_{\star}.
\label{eq:synch:x0}
\end{align}
With account of Equation (\ref{dotN_synchr}) the
thermalization condition (\ref{thermalization}) is now written as
\begin{align}
\frac{1}{3} \alpha^{-4/5}  \tauT^{1/5} \, \left[\frac{\epsinj}{\ginj (\epsB+\epsBB)}\right]^{4/5}
\left(\frac{E_B}{\kB T}\right)^{6/5} \theta^{-13/5} \ge 1.
\end{align}
Substituting $\theta$, $\tauT$ and $E_B$
from Equations (\ref{bb}), (\ref{eq:prob:tau}) and (\ref{eq:cycl:xB}),
we have finally
\begin{align}
0.8 \, \frac{\gamma_{{\rm acc},2}  \, r_{10}^{1/8} \, \Gamma_{1}^{3/8} \,
\Gamma_{{\rm max},3}^{1/4}}{L_{52}^{3/16}}
\le \frac{\epsB^{3/4}}{\epsBB^{13/16}} \, \frac{\epsinj}{\epsB+\epsBB}.
\label{eq:Synch:therm}
\end{align}
Note that the factor $\epsinj/(\epsB+\epsBB)$ in Equation (\ref{eq:Synch:therm})
is always smaller than unity because $\epsinj<\epsBB$, as $\epsBB$ includes
the total radiation energy while $\epsinj$ corresponds to the non-thermal dissipation channel only.
Taking into account a very weak dependence on $L$, $r$, $\Gamma$ and $\Gamma_{\rm max}$,
this condition in fact constrains the Lorentz factor of accelerated electrons from above.

The arguments leading to the above constraint are valid only if $E_0$ is below
the synchrotron energy for electrons with $\gamma=\ginj$, i.e. if
\begin{align}
E_0/E_B\le 0.3 \ginj^2.
\label{eq:synch:x0xs}
\end{align}
This results in a {\it lower} limit for $\ginj$, found by combining
Equations (\ref{eq:synch:x0}) and (\ref{eq:synch:x0xs}), along with
$\theta=\Epk/6\Gamma \me c^2$ and $\Epk$ from Equation (\ref{Amati}):
\begin{align}
\ginj \ge \, &10 \,\, \epsrad^{-1/6} \, L_{52}^{-5/12} \, r_{10}^{1/3} \,
\Gamma_{1}^{5/6} \, \Gamma_{{\rm max},3}^{1/6} \nonumber \\
&\times \epsB^{-1/12}  \left(\frac{\epsinj}{\epsB+\epsBB} \right)^{1/6}.
\label{eq:synch:gaccmin}
\end{align}
If condition (\ref{eq:synch:gaccmin}) is not satisfied, most of
the synchrotron photons are downscattered by the induced Compton process
and thermal Comptonization will be depleted of seed photons,
resulting in high-peaked spectra. In summary, the requirement that
$\Epk$ be kept below a certain value restricts $\ginj$ to a range
of values determined by Equations (\ref{eq:Synch:therm}) and (\ref{eq:synch:gaccmin}).

If complete thermalization is not achieved, the position of the spectral peak can be estimated
by equating the total dissipated luminosity to the number of upscattered
synchrotron photons times the energy they attain by saturated Comptonization:
\begin{align}
\epsBB L \approx \frac{16\pi}{3} \, c \, r^2 \, \Gamma^2 \, 3 \kB T\,   t_{\rm dyn}\dot{\cal N}_{\rm synch}.
\end{align}
This yields
\begin{align}
\Epk \approx \, &6\Gamma \, \kB T \, \frac{\epsrad}{\epsBB}
= 560 \,  \frac{L_{52}^{1/7} \, \Gamma_{1}^{5/7} \,  \Gamma_{{\rm max},3}^{1/7} \,
\gamma_{{\rm acc},2}^{4/7}}{\, r_{10}^{3/7} \,} \nonumber \\
&\times\frac{\epsrad}{ \epsBB^{2/7} \, \epsB^{3/7}}
\left(\frac{\epsinj}{\epsB+\epsBB} \right)^{-4/7} \mathrm{keV}.
\label{eq:synch:Epk}
\end{align}
In the domain of its applicability,
the theoretical scaling $\Epk \propto r^{-3/7}$ agrees well with the numerical results.
One can find a condition for the flow parameters by comparing
Equation (\ref{eq:synch:Epk}) to the observed $\Epk$ values from Equation (\ref{Amati}),
\begin{align}
3 \, \frac{\gamma_{{\rm acc},2}  \, \Gamma_{1}^{5/4} \, \Gamma_{{\rm max},3}^{1/4}}{L_{52}^{5/8}\, r_{10}^{3/4}}
\le \frac{\epsBB^{1/2}\, \epsB^{3/4}}{\epsrad^{7/8}} \, \frac{\epsinj}{\epsB+\epsBB}.
\label{eq:Synch:notherm}
\end{align}
Note that the radius appears in the denominator on the left hand side
of Equation (\ref{eq:Synch:notherm}), meaning that large radii are favoured
for producing low-peaked spectra. On the other hand, $r$ is still constrained
from above by the condition of saturated Comptonization, Equation (\ref{y>1}).

The synchrotron photons are complemented by cyclotron emission from thermal electrons.
To determine which is the dominant source of photons for Comptonization,
it is sufficient to compare the transition energies $E_0$ for the two processes,
given by equations (\ref{eq:cycl:x0}) and (\ref{eq:synch:x0}), respectively.
The process for which $E_0$ is larger, dominates.
We find that photons are supplied by synchrotron if
\begin{align}
2.3 \, \frac{r_{11} \, \Gamma_{1}^{11/7} \, \Gamma_{{\rm max},3}^{4/7}}{L_{52}^{11/14} \,
\gamma_{{\rm acc},3}^{4/7} \, \theta_{-2}^{11/7}} \ge \epsB^{3/14} \left( \frac{\epsB+\epsBB}{\epsinj}\right)^{4/7}.
\label{eq:Synch:CyclorSynch}
\end{align}

\begin{figure}[h]
\plotone{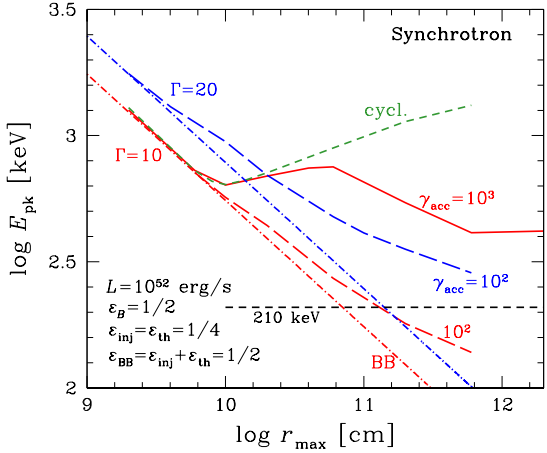}
\caption{The spectral peak energy as a function of dissipation radius for
complete and incomplete thermalization of synchrotron radiation.
The dissipation operates in the range $[\rmax/2,\rmax]$.
The solid and long-dashed lines correspond to the numerical results
with non-thermal particle injection with $\ginj=10^3$ and $10^2$, respectively.
The dashed green line is the numerical result with thermal heating only.
The dot-dashed lines correspond to the blackbody result (\ref{thermal_peak2}),
with $\epsrad=\epsBB=1/2$. The horizontal dashed line corresponds to
the average observed peak energy $210$ keV for $L_{{\rm rad},52}=\epsrad L_{52}=1/2$.}
\label{fig:Epk_synch}
\end{figure}

The behaviour of the peak energy with increasing dissipation radius is plotted
in Figure \ref{fig:Epk_synch}. The results with synchrotron emission from
non-thermal electrons are shown by solid and long-dashed lines for two
different electron acceleration Lorentz factors, $\ginj=10^3$ (solid)
and $10^2$ (long-dashed), and bulk Lorentz factors $\Gamma=10$ (red) and $20$ (blue).
High $\ginj$ results in few synchrotron photons and the radius where
thermalization fails is determined by cyclotron emission (solid line).
As the dissipation radius is increased, the peak energy initially behaves
as it would without non-thermal injection (short-dashed line). Synchrotron emission
becomes relevant near $r=10^{11}$ cm , where the condition (\ref{eq:Synch:CyclorSynch})
is fulfilled, and the increasing number of synchrotron photons leads to
a decrease of $\Epk$ with dissipation radius.

Lowering $\ginj$ increases the number of synchrotron photons. In the case
with $\ginj=10^2$ synchrotron emission becomes the dominant source
of soft photons approximately at the radius where complete thermalization fails.
Owing to Equation (\ref{eq:synch:Epk}) the peak energy only slightly deviates
from the blackbody case at larger radii (for which $E_{\rm pk,BB}\propto r^{-1/2}$).
The decrease of $\Epk$ continues as long as the Compton parameter is sufficiently large
to be able to upscatter most of the photons to $\sim 3\kB\Te$, i.e. Comptonization
proceeds in a saturated regime. The fact that the peak energy continues decreasing
with dissipation radius even after thermalization fails sets the synchrotron process
apart from all other mechanisms considered in this work. This results in the least
stringent constraints for the photon production site, which can extend to $r\sim 10^{12}$ cm.
Furthermore, full thermalization is no longer required for obtaining low-peaked spectra.
Nevertheless the peak energy compatible with the observations could not be obtained
at $\ginj\gtrsim 1000$, in agreement with the condition (\ref{eq:Synch:notherm}).

\begin{figure}[h]
\plotone{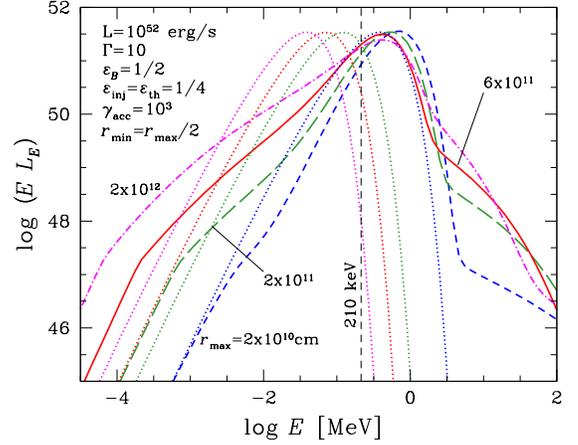}
\caption{Spectra at the end of dissipation, $\rmax$, from Comptonization
of synchrotron radiation. The short-dashed (blue), long-dashed (green),
solid (red) and dot-dashed (magenta) lines correspond to different dissipation radii.
The dotted lines are the corresponding Planck spectra for the same luminosities and radii.
The vertical dashed line corresponds to the average observed
peak energy $210$ keV for $L_{{\rm rad},52}=\epsrad L_{52}=1/2$.
}
\label{fig:Spec_synch}
\end{figure}

\begin{figure}[h]
\plotone{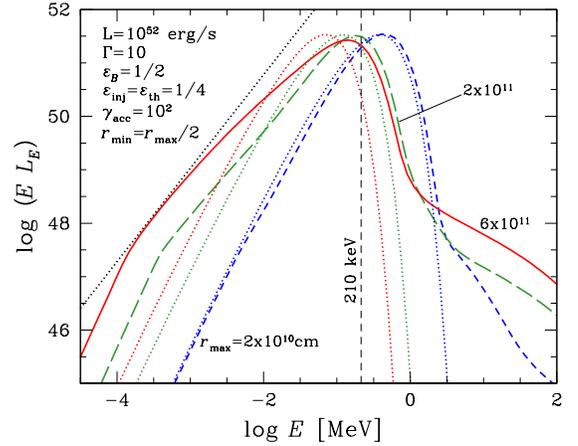}
\caption{Same as Figure \ref{fig:Spec_synch}, but for electron injection Lorentz factor $\ginj=10^2$.
The straight dotted line corresponds to $EL_{\rm E}\propto E^{7/4}$,
expected in the domain where the spectrum is determined by
the combined effect of synchrotron emission and downscattering
by the induced Compton process
(see Equation (\ref{ind_compt})).
}
\label{fig:Spec_synch:ginj1e2}
\end{figure}

\begin{figure}[h]
\plotone{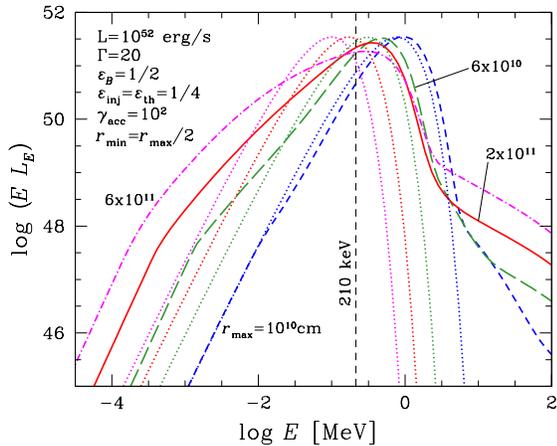}
\caption{Same as Figure \ref{fig:Spec_synch:ginj1e2}, but for bulk Lorentz factor $\Gamma=20$.}
\label{fig:Spec_synch:Gamma20}
\end{figure}

Figures \ref{fig:Spec_synch}, \ref{fig:Spec_synch:ginj1e2} and \ref{fig:Spec_synch:Gamma20} depict the spectra for
different injection energies and bulk Lorentz factors
$(\ginj,\Gamma)=(10^3,10)$, $(10^2,10)$ and $(10^2,20)$, respectively.
The different spectra in each figure correspond to different points on
the red solid and dashed lines in Figure (\ref{fig:Epk_synch}). In addition
to lowering the peak energy, increasing the dissipation radius leads to
progressive softening of the low-energy spectrum due to the decreasing
$y$-parameter. A broad range of low-energy slopes can be produced,
from thermal-like to almost flat in $E F_E$, covering nearly the entire
observed range. Furthermore, if $y$ is modest, the low-energy slopes
will survive up to the Thomson photosphere and directly correspond to
the observed spectra. Note, however, that if the dissipation radius is
too large so that $y\lesssim$ a few, the spectrum exhibits a broad maximum
and resembles neither the canonical Band shape nor a cutoff power-law
(e.g. the $r_{\rm max}=6\times 10^{11}$ cm case in Figure \ref{fig:Spec_synch:Gamma20}).

The observed high-energy emission
has to be produced at even larger radii, presumably close to $\tauT\sim1$.
The compactness parameter is very large in the photon-production region
and any radiation above $\me c^2$ (comoving frame) would be absorbed
by pair production. Also, as we require $y>1$, the analogous parameter
for downscattering $x \tauT > y >1$, thus photons above the spectral peak
are downscattered before they escape.

In the regime where the seed photons are provided by synchrotron,
two characteristic photon energies appear in the spectrum below the peak.
The higher of those, $E_0$, is the energy above which the produced photons
are up- rather than downscattered. The lower one is the energy below which
the emitted synchrotron photons are reabsorbed before they are significantly
downscattered by induced scattering. The spectrum between these frequencies
is determined by the combined effect of synchrotron emission and induced
downscattering and is given by Equation (\ref{ind_compt}); it
is compatible with the observed low-energy part of GRB spectra.

\section{Discussion}

In broad terms GRB emission models can be divided into two classes
based on whether the radiation is produced in optically thin regions
or near and below the Thomson photosphere. The most prominent
dissipation mechanism in the former category is the (collisionless)
internal shock model \citep{NPP92,ReesMeszaros94,SariPiran97,Kobayashi97},
while numerous models of energy release in Poynting-dominated jets
have also been considered
\citep[e.g.][]{Usov92,LyutikovBlandford03,Granot11,Lyutikov11,ZhangYan11,Granot12,McKinneyUzdensky12}.
The latter category includes various mechanisms invoking dissipation of either the magnetic field
\citep{Thompson94,Drenkhahn02,Thompson06,Thompson07,Giannios2008,Lyubarsky10}
or the kinetic energy of relative motions within the jet, e.g. via internal shocks
\citep{Peer2006,Bromberg11,Levinson12} or collisional dissipation \citep{B10,VBP11}.

A significant fraction of observed prompt GRB spectra
exhibit a low energy turnover that could not
be produced by radiatively efficient
optically thin emission.
Optically thick emission models avoid this problem, the local spectrum below
the photosphere can be as hard as Rayleigh-Jeans or even
Wien\footnote{The radiation received by the observer is subject to softening due to radiative transfer effects,
unless the flow remains radiation-dominated up to the Thomson photosphere (see e.g. \citealt{B10,B11}).}.
On the other hand, optically thick models face a different non-trivial issue:
the jet has to be able to generate a sufficient number of photons,
otherwise the spectrum peaks at energies that are too high compared with observations.
Our main finding in this work is that this requirement places severe constraints on the flow parameters
irrespective of the details of the subphotospheric dissipation mechanism.

Deep enough below the Thomson photosphere electrons and photons are in
equilibrium and share a common temperature (e.g. \citealt{Giannios12}),
the sufficient condition for which is that the Compton parameter $y\gg 1$.
However, even there the radiation does {\it not} necessarily have a Planck spectrum
since the interaction between photons and particles is mediated by
(single-) Compton scattering which conserves the number of photons.
A successful thermalization relies upon the existence of a sufficiently powerful
photon source, i.e. an emission/absorption process.

The importance of the thermalization issue lies in
its connection to the total number of photons carried by the jet at any given radius
and consequently the mean available energy per photon.
Consistency of the general thermodynamical relation between
the radiation temperature (i.e. the spectral peak position) and the energy density
with the observed $\Epk$ -- $L$ (Yonetoku) relation quite generally limits the thermalization region
to $r \gtrsim$ a few$\times 10^9$ cm if the flow is almost completely stopped at this radius,
and to even larger radii otherwise \citep{Eichler00,Thompson07}.
This is much larger than the size of the central engine, thus relic photons from
the center are insufficient to account for the observed spectral peak positions,
i.e. most of the observed photons have to be produced much further out.

We explored systematically all radiative processes capable of photon production in GRB jets:
 bremsstrahlung, double-Compton scattering, cyclotron and synchrotron emission.
In all cases efficient production of photons is confined to dense regions of the jet where
the Thomson optical depth is still very large. Thermalization can be achieved
either directly if the flow is optically thick to absorption or via saturated Comptonization
in which case the emission/absorption process acts as a source of soft photons.
The high density requirement sets an upper limit to the thermalization radius and,
most importantly, to the bulk Lorentz factor at the thermalization zone,
which is constrained to be at most $\Gamma\sim 10$
at radii $r\sim 10^{10}$--$10^{11}$ cm for bursts with typical luminosities and $\Epk$-s.
This rules out the possibility that
sufficient amount of photons could be produced in a freely expanding flow,
which would have acquired a substantially larger Lorentz factor by those radii.

Smooth unmagnetized jets yield the strongest constraints on the photon production
and/or thermalization conditions. For example, thermalization by bremsstrahlung
requires the jet to be at most mildly relativistic at radii $r \sim 10^{10}$ cm.
Under typical conditions within GRB outflows, photon production by bremsstrahlung is generally
superseded by double Compton scattering (see also \citealt{Beloborodov12}).
However, strong constraints for successful thermalization are obtained also
for this process: the Lorentz factor is limited to $\Gamma\lesssim$ a few
at the optimal radius $r \sim$ a few$\times 10^{10}$ cm and to even lower values at smaller or larger radii.

Strong local density enhancements/clumping
can make bremsstrahlung a plausible thermalization mechanism,
owing to the quadratic dependence of the rates on the density.
However, this requires regions of the flow to be compressed to
densities several orders of magnitude higher than the average.
The maximal conceivable compression can be expected in current sheets
in a Poynting dominated flow with an alternating magnetic field configuration,
which can yield $\sim 10^6$ for the ratio of densities in- and outside the current sheets.
We find that even in this case, consistency with the observed peak energies
can be achieved only if $\Gamma\le$ a few$\times 10$ at $r\sim 10^{11}$ cm.

In unmagnetized flows bremsstrahlung and double Compton are the only plausible
photon production mechanisms. In magnetized jets cyclotron and synchrotron emission
can also serve as photon sources. Thermalization by these processes relies on
saturated Comptonization to bring the emitted photons to the thermal peak.
This requires that the Compton parameter $y\gg 1$, which
restricts the Lorentz factor in the thermalization zone to at most $\sim 20$.
The constraints for generating sufficient amount of photons by cyclotron emission
are no less restrictive than those of bremsstrahlung or double Compton scattering.
Additionally, in this case one also requires a strong magnetic field.
In the optimal case of equipartition between magnetic and thermal energy,
one should have $\Gamma\lesssim$ a few near $r\sim 10^{10}$ cm to account
for the observed spectral peaks.

Finally, the observed photons may be produced by synchrotron emission
provided that a fraction of electrons can be accelerated to relativistic energies.
Although most of the emitted synchrotron photons are downscattered
by induced Compton owing to their high brightness temperature and subsequently reabsorbed,
the number of surviving/upscattered photons may be sufficient to account for observations.
In fact, the constraints for the dissipation radius and Lorentz factor for synchrotron,
$\Gamma\lesssim$ 10--20 at $r\sim 10^{11}$--$10^{12}$ cm,
are the least restrictive among all processes considered.
Most importantly, in this case a complete thermalization is not necessary for
obtaining low-peaked spectra. On the other hand, the mechanism requires
efficient channeling of power to acceleration of electrons, within a narrow range
of Lorentz factors, between $\ginj\sim 10$ and a few hundred, as well as a strong magnetic field.

\subsection{Possible thermalization models}

We turn now to examine several astrophysical processes that have been suggested as
sources of dissipation near the photosphere and examine whether the needed conditions for
photon production can be achieved in these scenarios.

The radius where a highly efficient
kinetic to internal energy conversion has to take place is in the vicinity of the
progenitor surface. Indeed, hydrodynamical simulations of the jet-star
interaction indicate that GRB outflows are far from simple passively expanding fireballs.
Instead, the outflow experiences strong heating via internal recollimation
shocks arising from interaction with the star as well as the hot cocoon
surrounding the jet \citep{ LazzatiBegelman05,BrombergLevinson07,Morsony07,Bromberg+11,LopezCamara12}.
\citet{Morsony07} find that after the jet head has broken out of the star,
 the subsequent portions of the outflow remain heavily disturbed as they emerge from
 the progenitor for up to tens of seconds. At this stage, the recollimation shocks
 occurring within the star keep the jet hot and Lorentz factor near the stellar surface
 relatively low ($\Gamma\sim 10$).
 This phase ends when the (first) recollimation shock reaches the progenitor surface,
 after which the unshocked, freely expanding jet emerges, with $\Gamma\sim 100$ at $r\sim 10^{10}$ cm.
Even under these favorable conditions the mildly relativistic Lorentz factors required
for efficient photon production and thermalization
in typical bursts are not attained.
However, the photon production rate can be sufficient
in the shocked jet phase for luminous bursts with higher $\Epk$-s, for which the constraint on $\Gamma$ is less stringent.

 At later times the recollimation shocks can still occur at progressively larger radii
 due to interaction of the jet with the
 expanding cocoon (\citealt{Lazzati09,Mizuta11}).
It was shown by \citet{Lazzati09} that heating by recollimation shocks can maintain a high
radiative efficiency all the way to the Thomson photosphere.
In addition to heating the flow these shocks slow it down and
the bulk Lorentz factor drops from a few hundreds to a few tens.
However, the detailed properties of the radiation field depend on the heating
history of the jet and subphotospheric dissipation doesn't automatically
give rise to a Planck spectrum. In the particular simulation shown by \citet{Lazzati09},
the recollimation shocks take place at $r\gtrsim 10^{12}$ cm and the Lorentz factor is
of the order of a few tens. Full thermalization is not successful under these conditions and
the spectrum is expected to peak at energies substantially higher than those typically observed.
Note also that the collimation shock is expected to be radiation mediated \citep{Bromberg11}
and shares the problems of these shocks discussed below.

An alternative possibility for efficient dissipation
has been proposed by \citet{GhiselliniCelotti07}. In this model the outflow is shocked by
encountering a small amount of slow material near the progenitor surface. For
example, shells in an intermittent outflow can collide with the progenitor
or cocoon material that has partially refilled the funnel excavated earlier by the jet.
A mass fraction $10^{-3}$--$10^{-4}$ of the displaced progenitor matter
is enough to have significant dynamical effects on the jet,
resulting in substantial conversion of kinetic energy to heat and in
slowing down the jet to $\Gamma\sim 10$.
Clearly, if such conditions can be achieved this
can plausibly result in complete thermalization of the dissipated energy.
However, the model is not sufficiently determined to check its consistency.
 An interesting point is that short GRBs show similar spectral features to long ones,
albeit with somewhat harder peak energies. These short GRBs most likely
don't involve collapsars \citep{Nakar07}. Therefore, both previously discussed
dissipation scenarios, which are based on collapsars, are not applicable to short GRBs
and one has to invoke a different mechanism to produce thermalization in these bursts.

The canonical internal shocks occurring in variable outflows can also give rise to
subphotospheric heating. Such shocks are expected to be mediated by
Compton scattering \citep{Levinson08,Budnik10,Katz10}, rather than plasma instabilities that operate in
collisionless shocks. The downstream pressure in radiation-mediated shocks in
GRBs is provided by photons advected from the upstream, the photon production
in the shock transition layer is negligible in comparison \citep{Bromberg11,Levinson12}.
The photon number density in the immediate downstream is substantially smaller than in the Planck spectrum,
which is compensated by a higher temperature than in the blackbody limit to support the required radiation pressure.
Thus the model is also subject to the photon production problem, leading to difficulty in producing low-peaked spectra.
Very recently, it was shown by \citet{Levinson12} that thermalization
further downstream is successful if the shocks take place at $\tauT\gtrsim 10^4$,
which is in general agreement with our results. However, since internal shocks are the result of
{\it relative} motions within the jet, it is unlikely that they can decelerate the flow
to $\Gamma\sim$ a few at $r\gtrsim 10^{10}$ cm necessary for obtaining low-peaked spectra.

We conclude that particularly for modest luminosity bursts, it is not easy
to find a subphotospheric dissipation model
that satisfies the conditions for sufficient photon production in unmagnetized jets. This
reflects the difficulty of slowing the jet down to sufficiently low
Lorentz factors at large enough radii.

The situation is even more speculative in the likely case where the flow is magnetized to some extent.
Shocks are unlikely to be radiatively efficient
even with a sub-dominant magnetic field \citep[e.g.][]{Narayan11}.
Furthermore, particle acceleration required for thermalization by synchrotron emission
is unlikely to work in shocks occurring deep in the flow irrespective of the flow magnetization.
In the weak field limit these shocks are radiation mediated and cannot accelerate particles to high
energies since the shock thickness is much larger than any kinetic scale
involved, which leads to particle thermalization rather than acceleration
\citep{Levinson12}.
It is also well known that particle acceleration even in moderately
magnetized shocks is not efficient \citep{Langdon88,SironiSpitkovsky2009,SironiSpitkovsky2011}.

It has long been speculated that the GRB jets are launched as Poynting dominated outflows.
Our results suggest that in this case, a significant part of the electromagnetic energy is released
at a scale of $10^{10}-10^{11}$ cm at relatively small Lorentz factors, much less than those
corresponding to the total conversion of the burst energy into the kinetic energy of the flow.
This could be achieved only via efficient magnetic dissipation, therefore one has to conclude that
the magnetic field is highly inhomogeneous at small enough scale so that
multiple current sheets are formed within the flow. There are two possibilities: a) the fields of alternating polarity
are imprinted already at the launch site so that a striped wind is formed; b) the initially regular magnetic field
is destroyed either by MHD instabilities \citep{LyutikovBlandford03} or by internal collisions \citep{ZhangYan11}.
The last mechanism could work only if the plasma inertia is not small, i.e. if the plasma kinetic energy is
a significant fraction of the magnetic energy,
whereas our upper limits on the flow Lorentz factor exclude this possibility in the thermalization region.
There is also evidence that MHD instabilities could hardly do the job; at least the kink instability,
which is the most dangerous one, is not disruptive in Poynting dominated flows \citep{McKinneyBlandford09,Mizuno12}.
Therefore the striped wind seems to remain the only viable option.

The striped wind structure arises naturally if the central engine is an obliquely rotating magnetar
\citep{Usov92,ZhangMeszaros01,Bucciantini07}. If the jet is launched by rotating, accreting black hole,
the structure of the magnetic field is less certain; possible scenarios for the formation of the striped wind
are discussed by \citet{Thompson94}, \citet{Spruit01} and \citet{McKinneyUzdensky12}.
An important point is that the magnetic dissipation is accompanied by the flow acceleration
so that most of the energy is released when the flow Lorentz factor is comparable with
the maximal one \citep{Drenkhahn02,DrenkhahnSpruit02,Lyubarsky10}.
However, a specific property of the magnetic dissipation mechanism is that
the acceleration is rather slow so that a significant fraction of the total energy
could be released when the flow has not been maximally accelerated
\citep{GianniosSpruit07,Giannios2008,Giannios12}.
It is also possible that within the progenitor star, the acceleration rate is
additionally reduced by interaction with the surrounding matter.
Therefore our findings do not exclude the striped wind scenario but place
severe limits that should be taken into account in future research.

\subsection{Conclusions}

In summary, we find that bulk of the observed GRB photons have to be produced
in the range $10^{10} \lesssim r \lesssim$ a few$\times 10^{11}$ cm,
at modest Lorentz factors of the order $\Gamma\sim 10$.
Since the available energy per baryon is typically much larger,
this means that the flow kinetic energy is small in that region,
and most of the available energy is in internal form (magnetic or thermal).
Photon production by processes involving only thermal electrons/pairs
yield the strongest constraints,
limiting the flow Lorentz factor to $\Gamma\lesssim 10$ in the thermalization zone for typical bursts,
and to only mildly relativistic values for low luminosity events.
The only exception is bremsstrahlung emission
from strongly clumpy flows, in which case the bulk Lorentz factor must still
satisfy $\Gamma\lesssim 30$. Synchrotron emission from relativistic electrons
can be a copious source of photons and relax the limit on the Lorentz factor
to $\Gamma\lesssim 20$.
However, it requires very specific conditions:
a highly efficient non-thermal acceleration mechanism transferring
a significant fraction of the total energy to electrons with Lorentz factors
between 10 and a few hundreds as well as a magnetic field near equipartition.

The constraints on $\Gamma$ are less severe for very luminous bursts
which typically exhibit higher spectral peak energies. The same is true
for outliers from the Yonetoku relation (towards higher $\Epk$ values),
however a 1$\sigma$ deviation relaxes the constraint on $\Gamma$ by at most a factor of $2$.
On the other hand, in this work we
demonstrated that there is a non-trivial problem of sufficient photon production
in typical GRB-s, a fact which does not depend crucially on the validity of the Yonetoku relation.

The constraints are also somewhat less restrictive if the radiation experiences
significant adiabatic losses on the way to the Thomson photosphere,
which might be the case if the flow reaches the coasting stage before transparency.
Adiabatic cooling lowers the mean photon energy, however this happens
at the expense of the radiative efficiency. Observationally, comparison of the energy
emitted in the prompt and afterglow phases indicates that a large fraction of
the available energy is typically released as prompt emission
(e.g. \citealt{Freedman01,Granot06,FanPiran06,Zhang07}), implying high efficiency.
There is also evidence of a positive correlation between the radiative efficiency
of the prompt phase and the burst luminosity (\citealt{LloydRonning04,Eichler05,Margutti13}).
The lower efficiency of less luminous lower-peaked bursts may alleviate
the photon-production problem which is otherwise more serious for low-luminosity bursts.

Finally, substantial decrease of the jet opening angle between the thermalization region
and the Thomson photosphere can also relax the constraints on the jet parameters
at the photon production site \citep{Beloborodov12}. Most of this range lies beyond
the progenitor surface where the opening angle is unlikely to change by a large factor,
thus the conclusions reached in this work should not be significantly affected.

Although in this work we addressed only the problem of the spectral peaks,
it is clear that the entire observed spectrum
and in particular the high-energy part cannot originate from the thermalization region.
At these radii the radiation compactness is huge and any high-energy photons
would be immediately absorbed by photon-photon pair production.
Furthermore, the high Compton parameter ensures that any photons above the thermal peak
will be downscattered even if they are below the pair-production threshold and can't be absorbed.
Thus the production of photons and their redistribution into non-thermal spectra must be the
result of either two different dissipation episodes or a single continuous one operating over
a very broad range of radii. Note in particular that (thermal) Comptonization that should
take place beyond the thermalization zone and produce the high energy tail cannot produce
photons at energies higher than $\sim \Gamma \me c^2$. Thus, the highest energy photons
must be produced by yet another, non-thermal mechanism giving rise to a relativistic
electron population, operating either in the prompt or the afterglow phase.

In baryonic jets, dissipation by neutron-proton collisions is a promising
mechanism for producing the high-energy tail \citep{B10}:
the requirement that bulk of the energy be in internal form at $r\sim 10^{10}$
along with the low $\Gamma$ facilitates the formation of a compound flow
(counter-streaming proton and neutron components). Compared to a passive fireball,
the saturation radius is now pushed further out relative to the neutron-proton decoupling radius,
i.e. the flow may still be accelerating when neutrons and protons start migrating
relative to each other, resulting in strong collisional heating. Internal shocks
can also provide additional energy release near the Thomson photosphere.

In Poynting-dominated jets, gradual dissipation of
the ``free'' magnetic energy associated with small-scale field reversals
can be responsible both for photon production
as well as distorting the spectrum into a
non-thermal shape.
Heating in current sheets maintains a thermal spectrum until the bremsstrahlung photosphere,
while continuing dissipation at larger radii
can give rise to the observed extended high-energy component
by Comptonizing the blackbody photons \citep{GianniosSpruit07,Giannios2008}.
Whether the mechanism can reproduce the observed spectral peaks depends on
the ability of the jet to keep its Lorentz factor low enough to sufficiently large radii.

\section*{Acknowledgements}

We thank Andrei M. Beloborodov for useful discussions that helped improve this manuscript
and for making his own manuscript available prior to submitting it for publication,
as well as Juri Poutanen for useful comments.
This research was supported by an Advanced ERC
grant and by the Israeli Center for Excellence for
High Energy Astrophysics,
and the Estonian Ministry of Education and Research Grant SF0060030S08 (IV).

\appendix
\section{Comptonization of soft photons}
\label{section:app:Compt}

In this Appendix, we address the evolution of the radiation spectrum in the comoving frame.
Well inside the scattering photosphere, the radiation is isotropic in the comoving frame
therefore the evolution of the photon spectrum is governed by the Kompaneets equation
\begin{align}
\frac{\partial n}{\partial t}=
\frac{\sigmaT N}{\me c} \frac 1{E^2}\frac{\partial}{\partial E}E^4\left(\kB T\frac{\partial n}{\partial E}+n+n^2\right)
+j(E)-\alpha(E)n+\frac 13(\mathbf{\nabla\cdot v})E\frac{\partial n}{\partial E},
\label{eq:meth:Komp}
\end{align}
where $n$ is the photon occupation number, $j(E)$ and $\alpha(E)$ are the photon emissivity and opacity,
respectively. The last term accounts for adiabatic cooling; in a spherical outflow with the Lorentz factor $\Gamma$,
\eqb
(\mathbf{\nabla\cdot v})=\frac{c}{r\Gamma}.
\eqe
In the Comptonization operator, the first term in the brackets describes redistribution
of soft, $E< \kB T$, photons towards $E\sim 3\kB T$, the second describes
the recoil effect negligible for soft photons and the third term is responsible for the induced scattering.

At the condition (\ref{Compton}), the Bose-Einstein spectrum with the peak
at $E=3\kB T$ is formed. At the same condition, one can neglect the adiabatic losses
(the last term in Equation (\ref{eq:meth:Komp})). Moreover, instead of solving
the full Equation (\ref{eq:meth:Komp})) one can now consider only the total photon balance,
 \eqb
 \frac{d\cal N}{dt}=\dot{\cal N},
 \eqe
where
\eqb
{\cal N}=\frac{8\pi}{(ch)^3}\int E^2 ndE
\eqe
is the total number of photons in the Bose-Einstein spectrum, $\dot{\cal N}$ is the rate
of injection of photons into the thermal peak. The latter is determined by the processes
at small photon energies, $E\ll \kB T$; by virtue of the condition (\ref{Compton})
the Kompaneets equation is reduced in this range to
 \eqb
-\frac{\sigmaT N}{\me c} \frac 1{E^2}\frac{\partial}{\partial E}E^4
\left(\kB T\frac{\partial n}{\partial E}+n^2\right)=j(E)-\alpha(E)n,
 \label{short_Kompaneets}\eqe
which means that the injected photons are steadily redistributed in energies
by the Comptonization process. The photon injection rate could be found from
the solution to this equation as the photon flux in the energy space
towards higher energies but one can find simple estimates just by
comparing different terms in this equation.

Note that the induced scattering (the $n^2$ term) dominates only if the radiation
brightness temperature, $T_{\rm b}=E n/\kB$, exceeds the electron temperature, $T$;
in this case the photons are redistributed towards {\it smaller} energies until
they are eventually absorbed. Such a situation arises if the synchrotron emission
serves as a photon source. If the emission/absorption processes are only due to
the thermal electrons, the thermodynamical condition, $j=(\kB T/E)\alpha$,
implies that the radiation brightness temperature does not exceed that of electrons.
In this case the induced scattering becomes only marginally important when
the spectrum approaches that of Rayleigh-Jeans so that for the estimates,
one can neglect the induced scattering at all and write Equation (\ref{short_Kompaneets}) as
 \eqb
-\frac{\sigmaT N\kB T}{\me c} \frac 1{E^2}\frac{\partial}{\partial E}E^4
\frac{\partial n}{\partial E}=\alpha(E)\left(\frac{\kB T}{E}-n\right).
 \label{short_Kompaneets1}\eqe
Since the opacity increases with decreasing photon energy, the right-hand side
of Equation (\ref{short_Kompaneets1}) dominates at small enough $E$;
the Rayleigh-Jeans spectrum is established in this band. At larger photon energies,
the Compton redistribution rate exceeds the absorption rate. The boundary energy, $E_0$,
is found by comparing the left-hand side and the right-hand side of equation (\ref{short_Kompaneets1}) as
\begin{align}
4\frac{\kB T}{\me c}\sigmaT N=\alpha (E_0).
\label{eq:tau_vs_y}
\end{align}
This relation could be also found from the following considerations. In order to avoid reabsorption,
an emitted photon has to experience significant upscattering before it is absorbed.
Thus if the Compton parameter $y$ (see equation (\ref{Compton})) associated with
the absorption time-scale, $t_{\rm abs}=1/\alpha (E)$, is greater than unity, the photon has
enough time to gain a significant energy and, since $\alpha$ decreases with $E$,
such a photon avoids reabsorption altogether and eventually reaches the thermal peak.
In the opposite case, the emitted photons are absorbed.
The condition (\ref{eq:tau_vs_y}) corresponds to the boundary, $y=1$.

Now let us consider specific mechanisms of the photon production.

\subsection{Bremsstrahlung}
\label{section:app:Brems}

At GRB conditions, bremsstrahlung can serve as a photon source only if
plasma is heavily loaded by electron-positron pairs (see Section \ref{section:brems})
therefore we use the bremsstrahlung opacity for the pair plasma,
which can be written as (e.g., \citealt{Pozdnyakov83,Haug85})
\begin{align}
\alpha_{\rm ff}^{\pm}(E)
&= \frac{\afs \, \gff(E,T)}{3^{1/2}  \,\pi^{3/2}}  \, \lambdaC^3 \,\sigma_{\rm T}cN_+N_-
\left(\frac{\kB T}{\me c^2}\right)^{-7/2}
\left(\frac{\kB T}E\right)^{3} \left( 1-e^{-E/\kB T}\right),
\label{eq:absff1}
\end{align}
or in the Rayleigh-Jeans regime
\begin{align}
\alpha_{\rm ff}^{\pm}(E)= \frac{\afs \, \gff(E,T)}{3^{1/2} \,\pi^{3/2}}  \, \left(\frac{\kB T}{\me c^2}\right)^{-7/2} \,
\left(\frac{\kB T}E\right)^{2}  \, \lambdaC^3 \,\sigma_{\rm T}cN_+N_-,
\label{eq:absff}
\end{align}
where $\lambdaC$ is the Compton wavelength and $N_{\pm}$ are
the electron/positron number densities. For the electron-proton plasma,
the opacity is just
\eqb
\alpha_{\rm ff}=\frac{N_{\rm p}}{2\sqrt{2}\,N_+}\alpha_{\rm ff}^{\pm},
\label{ep_ff}\eqe
where $N_{\rm p}$ is the proton density.
The Gaunt factor is approximated by \citep[e.g.][]{RL79}
\begin{align}
\gff(E,T) = \frac{\sqrt{3}}{\pi} \, \ln{\frac{2.35\kB T}{E}}.
\end{align}

From Equations (\ref{eq:tau_vs_y}) and (\ref{eq:absff}) we find the equation
for the boundary energy $E_0$ assuming $N_-=N_+$
\begin{align}
 \left(\frac{E_0}{\kB T}\right)^{2} = \frac{\afs \gff(E_0,T)}{3^{1/2} \, 8 \,\pi^{3/2}}
\left(\frac{\kB T}{\me c^2}\right)^{-9/2}N_-  \, \lambdaC^3.
\label{eq:nu0}
\end{align}
All the photons emitted at $E>E_0$ are redistributed to the thermal peak
so the photon injection rate is obtained as \citep{Illarionov75,Pozdnyakov83}
\eqb
\dot{\cal N}_{\rm ff}=\frac{8\pi}{(ch)^3}\int_{E_0}^{\kB T}E^2 j(E) \, dE=
4\pi\left(\frac{\kB T}{\me c^2\lambdaC}\right)^3\alpha_{\rm ff}^{\pm}(E=\kB T)\ln^2\frac{2.35\kB T}{E_0}.
 \label{Ndot_ff}\eqe

\subsection{Double Compton scattering}
\label{section:app:DC}

If the radiation energy density is large, double Compton scattering becomes
an important source of soft photons for Comptonization. The effective opacity
for a Maxwellian electrons interacting with a Wien distribution of photons
can be written at $E/\kB T \ll 1$ as \citep{Thorne81,Lightman81,Pozdnyakov83,Sve84}
\begin{align}
\alpha_{\rm DC}(E) &= \frac{2\afs}{\pi^2}  \left(\frac{\kB T}E\right)^{2} \theta^{-1} \gdc(\theta) \,
\lambdaC^3 \,\sigma_{\rm T}cN {\cal N},
\end{align}
where $\theta=\kB T/\me c^2 \ll 1$, $N$ and ${\cal N}$ are the electron and
photon number densities, respectively, and
\begin{align}
\gdc(\theta) = (1+13.91\theta+11.05\theta^2+19.92\theta^3)^{-1}
\end{align}
is a fitting formula to the exact numerical result.
For the Planck spectrum, one gets
\begin{align}
\alpha_{\rm DC}(E)
&= \frac{38.4\afs}{\pi} \left(\frac{\kB T}E\right)^{2} \theta^2 \gdc(\theta) \, \sigma_{\rm T}cN.
\end{align}

The calculation of the photon injection rate is similar to that for bremsstrahlung.
The produced photons are upscattered by single Compton scattering before reabsorption
above the energy $E_0$, found from Equation (\ref{eq:tau_vs_y}) as
\begin{align}
\left(\frac{E_0}{\kB T}\right)^{2} = \frac{9.6\afs}{\pi} \, \theta \gdc(\theta).
\label{eq:DC:nu0}
\end{align}
Now the photon injection rate is found as
\begin{align}
\dot{\cal N}_{\rm DC}
= 8\pi\left(\frac{\kB T}{\me c^2\lambdaC}\right)^3\alpha_{\rm DC}(E=\kB T)\ln\frac{\kB T}{E_0}.
\label{eq:DCinj}
\end{align}

\subsection{Thermal cyclotron emission}
\label{section:app:cycl}

The cyclotron radiation spectrum at high harmonics, $E\gg E_B$, could be considered
as continuous because the cyclotron line width becomes comparable with the line separation.
Since the electron temperatures relevant to our problem are of the order of $10$ keV,
we can use the Trubnikov approximation obtained for parameters appropriate for
thermonuclear fusion \citep{Trubnikov58},
\begin{align}
\alpha_{\rm cycl} = a \,\afs^{-1} \sigma_{\rm T}Nc \,\frac{\me c^2}{E}\, \theta^{q} \left( \frac{E_B}E \right)^{s}.
\label{eq:Synch:trub}
\end{align}
In \citet{Trubnikov58} the constants take the values $a=3 \, (120\pi)^2/2$, $q=3$ and $s=6$,
which provides a good fit at low harmonics.
At the energies relevant to the present problem, $E/E_B\sim10$,
the exact opacity already exhibits a significant downward curvature
and the Trubnikov approximation gives a poor fit.
We find that in the relevant parameter region
$a=10^{10}$, $q=4$ and $s=10$ provide a sufficiently accurate fit for our analytical estimates.

The cyclotron emissivity and absorption coefficient decrease very rapidly with frequency,
thus almost all the radiation is emitted near the energy $E_0$ defined by equation (\ref{eq:tau_vs_y}),
where the upscattering and reabsorption rates are equal.
Making use of the approximation (\ref{eq:Synch:trub}), one finds
\begin{align}
\frac{E_0}{E_B} = 14 \, \left(\frac{\me c^2}{E_B}\right)^{1/10} \theta^{3/10}.
\label{eq:cycl:x0}
\end{align}

At $E<E_0$, the radiation spectrum is Rayleigh-Jeans.
At $E>E_0$, the cyclotron emission and absorption could be neglected and
the spectrum is formed by Comptonization. The injection rate of photons into
Comptonization can be found from the Kompaneets equation (\ref{eq:meth:Komp})
by considering the photon flux in energy space at $E_0$,
\begin{align}
\dotN_{\rm cycl} = -\frac{4\pi  \sigmaT N\kB T}{\me c^4h^3} \, \left( E^4 \frac{\partial n}{\partial E}\right)_{E=E_0}
=\frac{12\pi\me}{h^3} \sigmaT N \theta^2 E_0^2.
\label{eq:cycl:dotN}
\end{align}
Here we substituted the photon occupation number as
\eqb
n=\frac{\kB T E_0^2}{E^3},
\eqe
because $n\propto E^{-3}$ is a solution to Equation (\ref{short_Kompaneets1}) when
the right-hand side is negligibly small; the normalization is chosen such that
the spectrum goes to the Rayleigh-Jeans at $E=E_0$.
Note that this solution describes a constant photon flux in
the energy space from $E=E_0$ to higher energies.

\subsection{Synchrotron emission}
\label{section:app:synch}

The synchrotron emissivity in GRB jets could be presented as (see discussion in Section \ref{section:synch})
 \eqb
j(E)=j_{\star}\left(\frac{E_B}{E}\right)^{7/2}.
 \eqe
The synchrotron absorption becomes important only at very low photon energies,
where the radiation brightness temperature is comparable with the energy
of emitting relativistic electrons. At the energies larger than the self-absorption energy,
the brightness temperature decreases but while it exceeds the temperature
of thermal electrons, the Comptonization is determined by the induced scattering therefore
the Kompaneets equation (\ref{short_Kompaneets}) is reduced to
 \eqb
-\frac{\sigmaT N}{\me c} \frac 1{E^2}\frac{\partial}{\partial E}E^4n^2=j(E).
 \eqe
The solution has a form \citep{Syunyaev71}
 \eqb
 E^3n=\left(\frac{2\me cE^5j(E)}{\sigma_{\rm T}N}\right)^{1/2}\propto E^{3/4}.
 \label{ind_compt}\eqe
Substituting this expression back into the Comptonization operator, one ensures that
the photon flux in the energy space is directed towards lower energies where
these photons are eventually absorbed.

The Compton upscattering begins when the brightness temperature decreases
down to the electron temperature. The boundary energy, $E_0$, may be found from
the condition that the two terms in the Comptonization operator
(the left-hand side of Equation (\ref{short_Kompaneets})) becomes equal.
For the spectrum (\ref{ind_compt}), this condition can be written as
\begin{align}
\frac{81}{32}\frac{\sigmaT N}{\me c} \frac{(\kB T)^2}{E_0}=j(E_0).
\label{eq:synch:upsc}
\end{align}
Note that if the emissivity were provided by the thermal electron population,
$j=(\kB T/E)\alpha$, Equation (\ref{eq:synch:upsc}) would be reduced,
to within a factor of about unity, to Equation (\ref{eq:tau_vs_y}) used when only
thermal electrons were considered. All photons emitted at $E>E_0$ are redistributed,
by virtue of the condition (\ref{Compton}), towards the thermal peak
therefore the photon injection rate is found as
\eqb
\dot{\cal N}_{\rm synch}=\frac{4\pi}{(ch)^3}\int_{E_0}^{\kB T}E^2j(E)dE=
\frac{8\pi E_0^3j(E_0)}{(ch)^3}.
\label{dotN_synchr}\eqe

\section{Numerical method}
\label{section:app:method}

The Kompaneets equation (\ref{eq:meth:Komp}) is coupled to the corresponding kinetic equation
governing the evolution of the electron distribution, which have to be solved simultaneously
for a self-consistent solution. However, in the present case the problem simplifies
owing to the fact that thermalization can only take place well below the Thomson photosphere
where bulk of the electrons are Maxwellian. Furthermore, the (thermal) lepton population has
negligible heat capacity and is therefore kept in a quasi-equilibrium temperature determined
by the combined effect of dissipation heating and radiative cooling. Thus solving a separate
kinetic equation for thermal electrons is not necessary. If the dominant cooling mechanism
is Compton scattering, the electron temperature is found by equating the volume heating rate
of electrons, $\Ph$, with the energy transfer rate determined by the Kompaneets kernel.
This yields
\begin{align}
\theta = \frac{\lambdaC^3\Ph/(8\pi\me c^3 \sigmaT N)+ \int x^4 \, n \, (1+n) \, \rmd x}{4\int x^3 \, n \, \rmd x},
\label{eq:meth:theta}
\end{align}
where $x=E/\me c^2$ is the dimensionless photon energy.
The volume heating rate $\Ph$ is determined by assuming
that a fraction $\varepsilon_{\rm BB}$ of the total available energy
is dissipated between $\rmin$ and $\rmax=\rmin + \Delta r$
from the central source.
For constant power per logarithmic radius interval, this gives
\begin{align}
\Ph = \frac{\epsBB L}{4\pi \rmin^3 \Gamma \ln(\rmax/\rmin)} \, \left( \frac{r}{\rmin} \right)^{-3}.
\end{align}
At each timestep, Equations (\ref{eq:meth:theta}) and (\ref{eq:prob:N})
together with the radiation field given by the Kompaneets equation (\ref{eq:meth:Komp})
determine the electron temperature and density and thus the entire thermal distribution.

If we allow a fraction of leptons to be accelerated to relativistic energies
(see Section \ref{section:synch} on thermalization by synchrotron emission),
we need a kinetic equation describing their cooling,
\begin{align}
\frac{\partial N(\gamma)}{\partial t} = -\frac{\partial}{\partial\gamma}
\left\{ \dot{\gamma} N(\gamma) + \frac{1}{2} \frac{\partial}{\partial\gamma} \left[  \De N(\gamma)\right] \right\}
+ \je - \alphae N(\gamma),
\label{eq:meth:elkin}
\end{align}
where $N(\gamma)$ is the electron distribution function, $\dot{\gamma}$ and $\De$ describe
the cooling/heating and diffusion in energy space due to synchrotron emission/absorption,
as well as Compton scattering in the Thomson regime. The source and sink
terms $\je$ and $\alphae$ account for Compton scatterings in the Klein-Nishina regime,
the former also includes the injection term accounting for accelerated electrons.

To solve the kinetic equations (\ref{eq:meth:Komp}) and (\ref{eq:meth:elkin}) we use
a numerical code developed by \citet{VP09} and \citet{VBP11}
for simulating radiative transfer in relativistic flows. For the present problem
the code has been modified in three main respects:
first, we have included induced Compton scattering which was not present
in the original version. This is crucial for obtaining true blackbody spectra (instead of Wien).
Furthermore, induced scattering dominates ordinary scattering in cases when
the radiation brightness temperature exceeds $\theta$, which can be relevant
if a suprathermal population of emitting leptons exists.
Secondly, we solve the equation for the angle-averaged photon distribution
instead of the full radiative transfer equation. This is justified by the huge
Thomson optical depth in the thermalization zone that keeps the radiation field
very nearly isotropic in the comoving frame. Finally, the electron equation is solved
only for the non-thermal population (when applicable). This avoids numerical difficulties
at low energies that are associated with keeping track of energy conservation
in near-equilibrium situations, where large cancelling terms can lead to errors
as well as numerical instabilities. These last two simplifications allow us to use
the familiar Kompaneets kernel in the photon equation, as well as to shorten
significantly the computation time.

\section{Constraint on thermalization location in a collimating flow}

\label{section:app:collim}

Consider a jet with a total (kinetic+internal, beamed) luminosity $\Lt$ and
beaming solid angle $\Delta\Omega(r)$ that may vary with radius.
The isotropic equivalent luminosity is now also a function of radius
and is given by $L(r)=4\pi\Lt/\Delta\Omega(r)$.
The peak energy of a blackbody spectrum is given by
\begin{align}
\Epk = 200\sqrt{\frac{\Gamma_2}{r_{12}}} \, \frac{\epsrad}{\epsBB}
\left(\frac{4\pi\,\epsBB  \, L_{{\rm t},52}}{\Delta\Omega(r)}\right)^{1/4}\,\rm keV.
\label{thermal_peak2:coll}
\end{align}
The Yonetoku relation now becomes
\begin{align}
\Epk=300 \left(\frac{4\pi\,\epsrad \,L_{{\rm t},52}}{\Delta\Omega(\rTh)} \right)^{1/2}\,\rm keV,
\label{Amati:coll}
\end{align}
where $\rTh$ is the radius where the radiation decouples from the flow
and gives rise to the observed emission. Equating the two expressions for $\Epk$,
we find the constraint for the thermalization location
\begin{align}
\frac{r}{\Gamma}=5\cdot 10^9 \, \frac{\epsrad}{\epsBB^{3/2}}  \,
L_{{\rm t},52}^{-1/2}\,  \frac{\Delta\Omega(\rTh)}{\left[ 4\pi\,\Delta\Omega(r)\right]^{1/2}} \, {\rm cm}
=5\cdot 10^9 \, \frac{\epsrad}{\epsBB^{3/2}}  \, L_{52}^{-1/2}(\rTh)\,
\left[\frac{\Delta\Omega(\rTh)}{ \Delta\Omega(r)}\right]^{1/2} \, \rm cm.
\label{r-gamma:coll}
\end{align}

Whether the photons from the central source are sufficient to account for
observed $\Epk$-s is determined by writing Equation (\ref{r-gamma:coll})
for the base of the flow, i.e. setting $r=r_0$ and $\Gamma=\Gamma_0\sim1$.
Assuming a quasi-isotropic flow at the base, $\Delta\Omega(r_0)=2\pi$,
we find the required opening angle of the jet at the photosphere
\begin{align}
\theta_{\star} \sim 3\times 10^{-3} \, \frac{\epsBB^{3/2}}{\epsrad}  \, r_{0,7} \, L_{52}^{1/2}(\rTh),
\label{eq:app:theta}
\end{align}
where we have used $\Delta\Omega=\pi \theta^2$.
Note that this also requires $\Gamma>1/\theta_{\star}$ since radiation
cannot be beamed into a cone narrower than $1/\Gamma$. The $\theta_{\star}$ values
imposed by Equation (\ref{eq:app:theta}) are at least an order of magnitude
below the typical opening angles inferred from afterglow observations \citep{Liang08,Cenko10}
strongly suggesting that the photons from the central engine are insufficient to
account for the observed spectral peaks and that most of the observed photons
must be produced in the jet at larger radii. Conversely, for a given opening angle
$\theta_{\star}$ the radius at the base of the flow has to satisfy
\begin{align}
r_0 = 3\times 10^8  \,\frac{\epsrad}{\epsBB^{3/2}}  \, L_{52}^{-1/2}(\rTh)\, \theta_{\star,-1} \, \rm cm,
\label{eq:collim:r0}
\end{align}
which is significantly larger than the size of the central engine.
Furthermore, the minimal value given by Equation (\ref{eq:collim:r0})
assumes that the flow is quasi-spherical at $r_0$.
This is implausible at radii indicated by Equation (\ref{eq:collim:r0}),
thus even if the flow is almost completely stopped at $r_0$, a more realistic
minimal radius for this to happen is $r_0\gtrsim 10^9$ cm.
Even more extreme constraints are obtained for magnetized jets,
in which case the factor $\epsBB$ relating the radiation luminosity to
the total luminosity at the base is much smaller than unity.

It is reasonable to expect that most of the flow collimation takes place
at relatively small radii in a dense environment providing strong confinement
for lateral expansion. At radii larger than the progenitor size, $r\gtrsim 10^{10}$ cm,
the angular factor in Equation (\ref{r-gamma:coll}) should not affect the constraint
on $r/\Gamma$ by more than a factor of a few.

\section{The impact of deviation from the Yonetoku relation on Lorentz factor constraints}

\label{section:app:yonet}

Let's rewrite the Yonetoku relation (\ref{Amati}) as
\eqb
\Epk=300 \, w \, L^{1/2}_{{\rm rad},52}\,\rm keV,
\label{Amati:dev}\eqe
introducing the factor $w$ to account for the scatter/deviation from the exact relation.
Following the arguments
leading to Equation (\ref{r-gamma})
in Section \ref{section:phprod},
the constraint on the thermalization location now becomes
\eqb
\frac{r}{\Gamma}=5\cdot 10^9 \, \frac{\epsrad}{\epsBB^{3/2}} \, w^{-2} \, L_{52}^{-1/2}\,\rm cm.
\label{r-gamma:dev}
\eqe
The constraint (\ref{saturated}) on $\Gamma$ from the requirement of saturated Comptonization changes to
\begin{align}
\Gamma \lesssim 21 \frac{\epsBB^{5/8} \, w^{3/4} L_{52}^{1/2}}{\epsrad^{3/8} \, \Gamma_{\rm max, 3}^{1/4}}.
\end{align}
The upper limits on the jet Lorentz factors at the thermalization location for different processes are modified as follows:
\begin{align}
\Gamma \lesssim 2 \, \frac{\epsBB}{\epsrad^{1/2}} \, w \, L_{52}^{1/2}, \qquad
\Gamma \lesssim 4.8 \, \frac{\epsBB^{7/10} w^{4/5} L_{52}^{1/2}}{\epsrad^{2/5} \, \Gamma_{{\rm max},3}^{1/5}}, \qquad
\Gamma \lesssim 5.4 \, \frac{\epsBB^{19/31}\epsB^{9/62} w^{26/31} L_{52}^{1/2}}{\epsrad^{13/31}\, \Gamma_{\rm max,3}^{5/31}}
\end{align}
for pair bremsstrahlung, double Compton and cyclotron emission, respectively
(cf. Equations (\ref{eq:brems:Gamma}), (\ref{eq:DC:Gamma}) and (\ref{eq:cycl:Gamma})).
For synchrotron emission, the requirement that the predicted $\Epk$ be compatible with the observed value
yields (cf. Equation (\ref{eq:Synch:notherm}))
\begin{align}
3 \, \frac{\gamma_{{\rm acc},2}  \, \Gamma_{1}^{5/4} \, \Gamma_{{\rm max},3}^{1/4}}{L_{52}^{5/8}\, r_{10}^{3/4}}
\le w^{7/4} \, \frac{\epsBB^{1/2}\, \epsB^{3/4}}{\epsrad^{7/8}} \, \frac{\epsinj}{\epsB+\epsBB}.
\end{align}

A 1$\sigma$ variation in the $\Epk$ -- $L$ relation towards higher values of $\Epk$ approximately corresponds to $w\approx 2$ \citep[e.g.][]{Yonetoku04,Nava08}, which can relax the upper limits on $\Gamma$ by about a factor of $2$.

\end{document}